\documentclass[11pt]{article}
\usepackage[dvips]{graphicx}
\parskip=\medskipamount

\newtheorem{prop}{Proposition} 
\newcommand{\BP}{\begin{prop}}   \newcommand{\EP}{\end{prop}}
\def\qed{$ \  \bullet$}

\def\be{\begin{equation}}
\def\ee{\end{equation}}
\def\bea{\begin{eqnarray}}
\def\eea{\end{eqnarray}}
\def\bean{\begin{eqnarray*}}
\def\eean{\end{eqnarray*}}
\def\MathR{\hbox {\rm I\kern -2.3pt  R}}
\def\MathC{\hbox {\rm {\tiny $\stackrel{\vert}{ \ }$}\kern -3.8pt  C}}

\def\eexp{ {\rm e  } }

\title{Observations on Gaussian Bases for Schr\"{o}dinger's Equation}
\author{Ilan Degani\footnote{email address: ilan.degani@mi.uib.no } \\
{\it Department of Mathematics, University of Bergen,} \\
{\it Johannes Brunsgate 12, 5008 Bergen, Norway  }}
\date{\today}

\begin{document}

\maketitle

\begin{abstract} 
One of the few methods for generating efficient function spaces for multi-D Schr\"odinger eigenproblems is given by Garashchuk and Light in {\it J.Chem.Phys. 114 (2001) 3929}. Their Gaussian basis functions are wider and sparser in high
potential regions, and narrower and denser in low ones. We suggest a modification of their approach based on the following observation: In very steep potential regions, wide, sparse, Gaussians should be avoided even if their centers have high potential values. 
Our numerical results
illustrate that a dramatic improvement in accuracy may be obtained in this way.
We also compare the errors of collocation to those of a Galerkin approach, test a criterion for scaling Gaussian widths based on deviation
from orthogonality of collocation eigenfunctions, and suggest a criterion for scaling Gaussian widths based on Hamiltonian trace minimization. 
\end{abstract}

\section{Introduction}  
Numerical prediction of the properties and behaviour of quantum mechanical
systems is a long standing challenge. A major difficulty stems from the high
dimensions of such systems. An $n \geq 3$ particle mechanical system generally has
$3 n - 6$ degrees of freedom, and so to calculate the state of an $n \geq 3$ particle quantum system we must solve a PDE in $3n - 6$ variables. Presently,
the limit is $n=5$ particles ($i.e.$ $9$ variables) in the largest systems for which many eigenvalues/eigenstates were obtained from Schr\"odinger's equation \cite{HighDimCalc}.
However, there are many larger systems for which numerical predictions would be
very useful for chemistry, biology (where many important molecules have more than five atoms), and for the emerging field of nano engineering. 
Clearly, we would like to push the limit further.

To efficiently treat multi-D ($i.e.$ multi variable) PDEs we must consider the function spaces where approximations are constructed. Naively, we can form such function spaces by taking tensor products of 1-D function spaces. 
Many of the 1-D spaces standardly used ($e.g.$ Fourier spaces, spaces of orthogonal
polynomials of different kinds) have a basis of localized functions peaked above the nodes of a 1-D quadrature formula, these are
called 1-D DVRs \cite{DVRoriginals}, \cite{LightDVR}.
The corresponding multi-D DVR basis of the product space consists of localized multi variable functions peaked above the points of ${\cal G}$, the (hyper) rectangular array of nodes of the associated product cubature formula. Typically,
at least $10$ 1-D DVR basis functions are taken for each variable, so the dimension of the product space is at least $10^{3n - 6}$. This gets much too large, even when the number of particles $n$ is not too big. 

DVR bases can give us some insight regarding the inefficiency of multi-D tensor product spaces. Bound solutions of Schr\"odinger's eigenproblem decay exponentially in high potential regions. Thus, for a given system we can apriori estimate $\Omega$, the region where wavefunctions up to a given energy (eigenvalue) are supported. Typically,
$\Omega$ is non rectangular and covering it with the rectangular grid ${\cal G}$
is inefficient. Many nodes will be outside $\Omega$ and therefore many of the multi-D product DVR basis functions will contribute little to the approximation of our wavefunctions. Evidently, efficient localized basis functions should be peaked in $\Omega$, and not outside it. Several approaches have been suggested for obtaining such basis functions.

One very useful solution is to prune a
product DVR basis by discarding functions whose nodes lie in high
potential regions \cite{LightDVR}. Another approach, still in its initial 
stages, is to construct DVRs based on non product cubatures
with various non rectangular node distributions \cite{NonProdDVR}. 
In both cases the resulting DVR basis functions are orthogonal. However, in these approaches it is not easy
to refine the node distribution $within$ $\Omega$ according to predictable wavefunction properties. 
Generally, wavefunctions are more oscillatory in low
potential regions compared to high ones and node densities in $\Omega$ should vary accordingly. Therefore we may try to construct function spaces by appropriately
distributing nodes and then placing a localized basis
function above each. Gaussians \cite{HamLight} \cite{PoirLight} \cite{GarLight}, B-splines \cite{BSplines}, and inverse multi quadrics \cite{RabitzGIMQ}, have been used in this way. 
Although the resulting basis functions are not orthogonal, the flexible choice of nodes and basis function shape parameters can give highly accurate results. This has
been illustrated in 1-D problems \cite{HamLight} \cite{PoirLight}. However the only general method we are aware of for
constructing multi-D bases of localized functions whose node density varies with potential values is given by Garashchuk and Light \cite{GarLight}.

Here we closely examine the method of \cite{GarLight} for choosing nodes and
widths of Gaussian basis functions. Their basis functions are
narrow and close together in low potential regions, while they are wide and sparsely distributed in high potential regions. We believe that this needs to be refined in
steep potential regions, where points with low and high potential values are very close together. This situation is common in molecular potential surfaces for small nuclear separation. We suggest a simple modification of the
nodes/widths choice of \cite{GarLight}. Its essence is to avoid
overly wide and sparse basis functions in hard potential regions. We illustrate that this can give a dramatic improvement in accuracy by calculating the eigenfucntions and eigenvalues of two 1-D Morse Hamiltonians. 
Our results indicate that the algorithm of \cite{GarLight} should be adapted along these lines in the multi-D case. 

Our work also explores the use of collocation with Gaussian
basis functions. Using collocation, rather than the standard Galerkin approach, is 
very appealing since calculating integrals of potential matrix elements is
avoided. Yang and Peet have previously used collocation with Gaussian basis
functions \cite{YangPeet1}, and in their calculations performance was similiar to the Galerkin approach. They have used the same basis functions with both methods.
However in our calculations basis function widths were adjusted optimally for each method and Galerkin accuracy was much better than collocation. Collocation has many advantages and it can give high quality results as shown in the work of Yang and Peet \cite{YangPeet1}, \cite{YangPeet2}, \cite{YangPeet3}, and in many papers based on it. However, our results illustrate that caution should be exercised before embracing collocation as equally accurate to Galerkin. 

Other aspects we examine are the use of collocation with the quasi randomly
distributed Gaussians of \cite{GarLight}, and using the deviation from orthogonality
of eigenfunctions calculated by collocation to guide us in the choice of basis function widths.

The content of this paper follows. In section \ref{DistGaussBases} we describe the distributed Gaussian bases used for Schrodinger's equation leading up to \cite{GarLight}, then we overview the collocation approach.
In section \ref{ObsSug} we discuss the nodes/widths choice of \cite{GarLight}, suggest improvements, and raise a few further questions about collocation and Gaussian basis
functions. Section \ref{NumEx} gives numerical results for two 1-D Morse Hamiltonians. Finally, we conclude and suggest some future directions in section \ref{Conclude}.

\section{Distributed Gaussian bases and collocation}     
\label{DistGaussBases}
Here we discuss the application of distributed Gaussian bases and collocation in the Schr\"{o}dinger eigenproblem, $i.e.$ the problem of finding eigenvalues and eigenfunctions of the Hamiltonian operator
\be
  \hat{H} = - \frac{\hbar^2}{2 m} \Delta + V(x) \ . \label{TISE}
\ee
We concentrate on the problem of finding all bound states below an energy cutoff $E_{max}$, $i.e.$ all eigenfunctions $\{ \psi_n \}$ and eigenvalues $\{ E_n \}$ belonging to the discrete part of the spectrum with $E_n \leq E_{max}$. 

Several researchers have studied the use of Gaussian basis functions for this problem\footnote{A thorough discussion of other types of basis functions is beyond the scope of this work. However, the performance of well chosen Gaussian bases is {\it good}. In \cite{BSplines} Galerkin b-spline calculations for Morse Hamiltonians are reported. Generally, results obtained there with approximately $100$ B-splines have much greater errors
compared to results obtained with $48$ Gaussians in \cite{HamLight},
\cite{GarLight}, and in section \ref{NumEx} here (to be fair we note that the Morse Hamiltonians were not identical). Moreover, it is illustrated
numerically in the first reference of \cite{BSplines} that eigenvalue errors using $n$ Gaussians scale as $\alpha^{-n}$ while the error using $n$ B-splines of order $k$ scales as
$\left( \frac{1}{n} \right)^{2k}$, $i.e.$ the Gaussian error converges asymptotically faster. A thorough analytical and numerical investigation of using high order B-splines in Schr\"odingers equation is certainly in place. However, we presently opt for Gaussians because of
the reasons given above, and because the lessons on node/width distributions we obtain using Gaussians will probably be useful for other types of basis functions including B-splines.
}.
We pick up an important thread in this research by considering the work of Hamilton and Light \cite{HamLight}, and continuing to Poirier and Light \cite{PoirLight}, and to Garashchuk and Light \cite{GarLight}. 

The basis functions used in \cite{HamLight} are real Gaussians
\be  \varphi_i(x) = \eta_i \eexp^{-c w_i (x - x_i)^2} \qquad i=1,\ldots, n 
\label{GausBasisF} \ee  
whose parameters are named as follows.
\begin{itemize}
\item $x_i$ is called the center or node of $\varphi_i$.
\item $w_i > 0$ is the width parameter of $\varphi_i$.
\item $c > 0$ is the ``global'' width parameter.
\item The normalization parameter $\eta_i$ is chosen so that $\langle \varphi_i | \varphi_i \rangle = 1$.
\end{itemize}
Much depends, of
course, on the choice of the $2 n + 1$ parameters $c$, $\{ w_i \}$, $\{ x_i \}$. For 1-D problems classical mechanics arguments are used in \cite{HamLight} to construct very efficient bases, ``tailored'' to a given problem. However, such arguments are not readily extended to multi-D, and a simpler approach is applied in \cite{HamLight} in this case. Its rationale can be explained as follows.
The highest eigenfunction with energy (approximately) $E_{max}$ is the most oscillatory in low potential regions, while reaching furthest out to high potential regions. Therefore a basis suitable for its approximation will be suitable for all lower eigenfunctions. 
The most severe oscillations appear near the potential minimum where the local De-Broglie wavelength is
$2 \pi / \sqrt{2m (E_{max} - V_{min})/\hbar^2}$. Thus the $\{ x_i \}$ are chosen in \cite{HamLight} as points of a regular grid whose spacing is at most $1/4 \pi$ of this De-Broglie wavelength (any sufficiently fine spacing would do). 
On the other hand, the highest eigenfunction is exponentially decaying in the classically forbidden region where $V(x) > E_{max}$, thus the $x_i$ are confined to the classically 
accessible region $V(x) < E_{max}$. Regular grids are used in \cite{HamLight} which allow an easy notion of grid spacing, however
this is by no means a restriction to rectangular grids.
For the Henon-Heiles problem the $x_i$ chosen in \cite{HamLight} are points
of a regular hexagonal grid pruned to fit within the classically accessible  triangular region.
Regarding width parameters,
the choice $w_i = 1/{\Delta x}^2$, is advocated in \cite{HamLight}. It is interesting to note that this choice arises if we require the curvature of the $\varphi_i$ at their centers $x_i$ ($i.e.$ $\frac{d^2 \varphi_i}{d x^2} |_{x=x_i}$) to be compatible with the maximal curvature of the highest eigenfunction. 
Having chosen the $w_i$ we can still vary the global width parameter $c$. 
Let us define the matrix $S$ by
\be S_{ij} = \langle \phi_i | \phi_j \rangle \ . \ee
In \cite{HamLight} the most accurate Hamiltonian eigenvalues are obtained for low $c$ values for which the condition number of
$S$ is large while still allowing stable numerical calculations.

Hamilton and Light \cite{HamLight} provide efficient Gaussian bases whose centers occupy only relevant regions of configuration space. However, for
multi-D problems the density of centers in \cite{HamLight} is constant, and dictated by the highest oscillations in the problem. Such high center density is necessary only in the lowest potential regions. As in the 1-D examples in \cite{HamLight} we expect that sparser Gaussian centers can be taken in higher potential regions. This is the subject of \cite{GarLight}, but we first discuss 
\cite{PoirLight} which provides the background for \cite{GarLight}, and which unveils important, and surprising, properties of Gaussian bases.

Intuitively, we tend to reject badly conditioned basis functions which are ``almost linearly dependent''. Thus we may expect that ``good'' Gaussian basis functions will not be ``too wide'' and their centers will not be ``too close'' together. In \cite{PoirLight} Poirier and Light show the contrary; that ``good'' Gaussian bases can actually have clustered centers and high condition numbers. They regard the choice of Gaussian basis parameters as an 
optimization problem.
Let us define the matrix $H$ by
\be H_{ij} = \langle \varphi_i | \hat{H} \varphi_j \rangle \ . \ee
Poirier and Light solve for widths and centers which minimize the functional $trace(S^{-1} H)$. 
This is based on the ``variational principle'': each eigenvalue of $\hat{H}$ is smaller than the corresponding eigenvalue of $S^{-1} H$, see MacDonald \cite{MacDonald}. Therefore minimizing
$trace(S^{-1} H)$ is equivalent to minimizing eigenvalue errors. 
Note that $S^{-1} H$ represents $\hat{H}$ as a {\it linear operator} in the non-orthogonal Gaussian basis, while $H$ represents the bilinear form $\langle \cdot | \hat{H} \cdot \rangle$. Note also that
including $S^{-1}$ in our functional we ensure that the Gaussians yielded by the optimization will be linearly independent, however their condition number is not constrained. 

An illumating example is provided by the 1-D harmonic oscillator, $\hat{H} = - \frac{\hbar^2}{2 m} \frac{\partial^2}{\partial x^2} + \frac{1}{2} m \omega^2 x^2$. Rather than giving a ``well conditioned'' basis, the optimization produces Gaussians which are small perturbations of the harmonic oscillator ground state
$\psi_0(x) = k_0 \eexp^{-\frac{\alpha^2}{2} x^2}$, where $\alpha = \sqrt{m \omega / \hbar}$.
These basis functions have large overlaps, and it is reported in \cite{PoirLight} that they reproduce the harmonic oscillator eigenvalues to machine accuracy. At first this may seem surprising, but a simple explanation is provided in \cite{PoirLight}. The $n$'th harmonic oscillator eigenstate is
$\psi_n(x) = k_n \eexp^{-\frac{\alpha^2}{2} x^2} H_n(\alpha x)$, where $H_n$ is the $n$th Hermite polynomial. Using the formula 
$\eexp^{- y^2} H_n(y) = (-1)^n \frac{d^n}{ d y^n} \eexp^{- y^2}$ (from $e.g.$ \cite{Arfken}) we see that $\frac{d^n}{ d x^n} \eexp^{- \frac{\alpha^2}{2} x^2} = \left( \frac{-\alpha}{\sqrt{2}} \right)^n \eexp^{- \frac{\alpha^2}{2} x^2} H_n(\frac{\alpha}{\sqrt{2}} x)$, and therefore $\psi_n$ is in
$span\{ \frac{d^k }{ d x^k} \eexp^{- \frac{\alpha^2}{2} x^2} | k = 0,\ldots,n \}$. Consider now the space mentioned above of Gaussians closely clustered near the potential minimum. It contains difference approximations to the derivatives  
$\frac{d^k }{ d x^k} \eexp^{- \frac{\alpha^2}{2} x^2}$, and since the centers are close together the accuracy of these difference approximations is high. Therefore $n$ closely clustered translations of the ground state span a linear space which gives excellent approximations of the first $n$ harmonic oscillator eigenstates. Clustering of optimal Gaussian centers is observed 
in \cite{PoirLight} also for a Morse potential, however in this case there are several clustering locations away from the potential minimum. The high condition number of such optimal Gaussian bases can be problematic in practical large scale calculations. However, the harmonic oscillator example indicates that high overlaps are needed only to produce accurate approximations of Gaussian derivatives. Poirier and Light therefore propose to include derivatives of Gaussians among our basis functions to begin with, and so avoid the need for large Gaussian overlaps. This direction is not pursued further in \cite{PoirLight}. Instead, the bad conditioning of optimal Gaussian bases is addressed by introducing a modified functional, \be trace(S^{-1} H) + \lambda \|S-I\| \ . \label{FuncPL} \ee
Gaussians optimizing this functional strike a compromise, depending on the value of $\lambda$, between accuracy and conditioning. A similiar functional provides the starting point for \cite{GarLight}.  

In \cite{GarLight} Garashchuk and Light suggest a practical way of constructing multi-D Gaussian bases,
whose node density varies with potential values. They define the functional
\be trace(H) + \lambda \sum_{i \neq j} \frac{S_{ij}}{1-S_{ij}} \ , \label{FuncGL} \ee 
which is different from the functional (\ref{FuncPL}) used in \cite{PoirLight}.
Note particularly that $trace(S^{-1} H)$ from (\ref{FuncPL}) is replaced by
$trace(H)$ in (\ref{FuncGL}).
 This is minimized for 1-D Morse and double well Hamiltonians to find the density of optimal Gaussian centers $\rho(x_i) = \frac{1}{2(x_{i+1} - x_{i-1})}$ and their widths $w_i$. It is then observed that $\rho(x_i)$ and $w_i$ fit linear functions of the potential, $i.e.$ there are constants $a_0, a_1, b_0, b_1$ such that $\rho(x_i) \approx a_0(a_1 - V(x_i))$,
and $w_i \approx b_0(b_1 - V(x_i))$. A similiar optimization for realistic large scale problems would be very time consuming. However, Garashchuk and Light regard these two 1-D examples as manifestations of a general rule: ``Gaussian bases which have center density and width parameters fitting appropriate linear functions of the potential will be good''. It is also observed in \cite{GarLight} that Hamiltonian eigenvalue errors decrease when the global width parameter $c$ is decreased until unstable numerical calculations set in when $cond(S) > 10^{12}$.
Based on these observations a simple procedure for choosing multi-D Gaussian basis functions is suggested in \cite{GarLight}. Fixing $E_{max}$ and a parameter $\mu$
they define the density function
\be \rho(x) = \left( \frac{E_{max} - V(x) + \mu}{E_{max} + \mu} \right)^\gamma \ . \label{DensityGL} \ee
With $\gamma = 1$ the density $\rho$ depends linearly on the potential. The
value $\gamma = 1/2$ is also considered in \cite{GarLight}, in this case $\rho$
corresponds to the local De-Broglie wave length of the highest eigenfunction.
Having defined $\rho(x)$ candidate points $x$ are quasi-randomly generated in the region $V(x) < E_{max}$. 
$x$ is accepted as a Gaussian center if
\be \rho(x)  > r  \ee
where $r \in [0,1]$ is random.
With the centers at hand, the widths are defined
\be w_i = E_{max} - V(x_i) + \mu > 0 \ . \label{WidthsGL} \ee 
Then $c$ is taken as small as possible until $cond(S) \approx 10^{12}$.  

This approach is applied in \cite{GarLight} to eigenvalue calculations of
several multi-D Hamiltonians. The results are much better, or not
worse, compared to those of previous methods. Therefore the quasi random distributed Gaussians of \cite{GarLight} appear to be a very good choice. 
It is our purpose here to make some further observations and suggest a few possible refinements of the approach of \cite{GarLight}. Among other things we shall use the Gaussian basis functions of \cite{GarLight} in a collocation algorithm, which we discuss next.

The collocation approach for finding the eigenvalues and eigenfunctions of a
Hamiltonian $\hat{H}$ may be described as follows. First, we choose 
basis functions $\varphi_1,\ldots,\varphi_n$, and a set of points $x_1,\ldots,x_n$ called collocation nodes. We then seek
functions $\psi = \sum_{i=1}^n u_i \varphi_i$, and eigenvalues $E$, such that
$(\hat{H} \psi)(x_i) = E \psi(x_i)$, $i=1,\ldots,n$.
This gives the following generalized matrix eigenproblem
\be \left(- \frac{\hbar^2}{2 m} \Delta \Phi + V \Phi \right) u = E \Phi u ,
\label{CollGeq} 
\ee
where $\Phi_{ij} = \varphi_j(x_i)$,
$(\Delta \Phi)_{ij} = \Delta \varphi_j(x_i)$, $V_{ij} = \delta_{ij} V(x_i)$.
Since the exact eigenvalues are real we reject solutions with complex $E$, if there are any.
Assuming $\Phi$ is invertible (\ref{CollGeq}) is equivalent to
\be  \left(- \frac{\hbar^2}{2 m} \Phi^{-1} (\Delta \Phi) + \Phi^{-1} V \Phi \right) u = E u  \label{CollCeq}
\ee
and changing coordinates to $y = \Phi u$ we obtain
\be \left(- \frac{\hbar^2}{2 m} (\Delta \Phi) \Phi^{-1} + V \right) y = E y
\ . \label{CollYeq}                                                    
\ee
The matrix on the l.h.s. of (\ref{CollCeq}) is the collocation representation
of the Hamiltonian in the $\varphi_i$ basis, while the matrix on the l.h.s. 
of (\ref{CollYeq}) is the collocation representation in a basis $\theta_i$,
spanning the same function space, and consisting of functions which are $1$ on ``their'' node and
zero on all other nodes, $\theta_i(x_j) = \delta_{ij}$. 
                                
The collocation approach for spatial discretization of differential 
operators has a long history \cite{InitialCollRefs}. An important mile stone
was the introduction of radial basis function (RBF) collocation by Kansa
\cite{KansaEtc}, an approach which found much use in the PDEs of classical physics.
For recent reviews on radial basis functions and their applications see \cite{MathRadBFuncs}.
RBFs have the form $\varphi_i(x,w_i) = f(\| x - x_i \|,w_i)$, where $f$ is a 
real valued function peaked at $0$ and depending on the parameter vector $w$,
$e.g.$ $f(r,w) = \eexp^{- w r^2}$ giving 
Gaussian RBFs, or $f(r,[w_\alpha,w_\beta]) = (1 + w_\alpha^2 r^2)^{-w_\beta/2}$ giving generalized inverse multi quadric (GIMQ) RBFs
(here $w_\beta > d$ where $x \in \MathR^d$). The centers $x_i$ are usually chosen as the collocation nodes. 
The attraction of RBF collocation stems from its great flexibility and simplicity. The basis functions shape and location can be adapted to a given problem by carefully choosing parameters. Moreover, the collocation equations are very simple to set up, they do not require calculation of multi dimensional integrals. 

The need for efficient methods is particularly acute for high dimensional
problems in quantum mechanics. Yang and Peet were the first to show that collocation is feasible for the Schr\"odinger eigenproblem  \cite{YangPeet1}. They applied collocation to 1-D Morse Hamiltonians together with the Gaussian RBFs of \cite{HamLight}, and found that results were as good as the Galerkin approach with the same basis functions. The RBF collocation of \cite{YangPeet1} has become a standard method for 1-D Schr\"odinger eigenproblems (appearing as components in more complicated calculations). In multi-D problems collocation did not go, to
the best of our knowledge, beyond the stage of initial explorations.
Collocation was used for a 2-D problem (Ar-HCl Hamiltonian) in \cite{YangPeet2} with rectangular grids of nodes and tensor product function spaces. Again, performance was similiar to that of a Galerkin approach with the same basis functions. In \cite{YangPeet3} collocation was used for the same 2-D problem with Gaussian RBFs whose centers were obtained by truncation of a rectangular grid outside the classically accessible region. However, in \cite{YangPeet3} only the first three eigenvalues were calculated using an iterative scheme based on projections to low dimensional subspaces. 
The only example we are aware of for calculation of high (additionally to low) lying eigenvalues and eigenfunctions of multi-D Hamiltonians using RBF collocation is given by Hu, Ho, and Rabitz \cite{RabitzGIMQ}. The nodes used in \cite{RabitzGIMQ} were on hexagonal and triangular grids and the basis functions were the GIMQ RBFs mentioned above. No direct
comparison with the Galerkin approach was given in \cite{RabitzGIMQ}.

Having surveyed the topics of Gaussian RBFs and collocation for the Schr\"odinger eigenproblem we can now venture to take a closer look at some aspects of these topics.



\section{Observations and suggestions}
\label{ObsSug}
Here we give some observations and suggestions which arose while studying
the application of Gaussian RBF collocation to Schr\"odinger equations. 
Although the term RBFs is used below for Gaussians, some
of our observations may be relevant also for other types of RBFs.

\begin{enumerate}
\item One of the few methods for constructing efficient function spaces for
multi-D quantum problems is given by the quasi random distributed Gaussians of \cite{GarLight}. However, the choice made in \cite{GarLight} for the width parameter of an RBF $\varphi_i$ depends only on the potential value at {\it a point}, see equation
\ref{WidthsGL}, 
while $\varphi_i$ should capture solution properties in {\it a neighborhood} of $x_i$. Such
properties correspond to the differing nature of the potential function in various regions. Strong repulsions cause molecular potentials to rise very steeply when distance coordinates get small. On the other hand, the smooth interaction of distant particles causes molecular potentials to be much softer for large values of distance coordinates. See for example the Morse potential in figure \ref{fig1}, which is a standard model for two atom molecular potentials. 

A possible problem in the width choice of \cite{GarLight} can be illustrated by considering two RBFs $\varphi_i$ and $\varphi_j$ such that $V(x_i) = V(x_j)$ are close to the cut off value $E_{max}$, and such that $x_i$ is in the ``hard'' region of the Morse potential while $x_j$ is in the ``soft'' one. Our assumptions imply that $\varphi_i$ and $\varphi_j$ will have the same large width. In the soft region this is fine, wide basis
functions are compatible with the smooth character of wavefunctions in this region. However
a very wide basis function seems unsuitable in the hard region where wavefunction behaviour can change from severe oscillation to rapid decay
over a short distance. We therefore conclude that RBFs centered in hard potential regions should generally be narrower than those centered in soft potential regions. 
This observation is supported by \cite{PoirLight} where widths and centers of $optimal$ Gaussian RBFs (w.r.t. functional (\ref{FuncPL})) are found for a Morse potential. Comparing centers with similiar potential values we see that the width for a center in the hard region is smaller than for a center in the soft region
(see particularly figure 2 in \cite{PoirLight}). Recall that functional
(\ref{FuncPL}) is different from functional (\ref{FuncGL}) used for optimizing RBF parameters
in \cite{GarLight}.
While the reasoning giving (\ref{FuncPL}) is clear, we find no similiar justification for
(\ref{FuncGL}). Particularly, it is not clear why the representation of $\hat{H}$ as a bilinear
form, rather than a linear operator, appears in (\ref{FuncGL}).
 
We suggest that the parameter choice of \cite{GarLight} may be emiliorated
by appropriately decreasing the widths of RBFs, and increasing 
their density, in hard potential regions. One way to achieve this is to choose nodes and width parameters as in \cite{GarLight} but using a modified potential. Let us denote the classically accessible region by $\Omega_c$,
the ``soft'' potential region by $\Omega_s$, and the ``hard'' potential
region by $\Omega_h$. 
We define $\tilde{V}(x)$ such that $\tilde{V}(x) = V(x)$ for
$x \in \Omega_s \cap \Omega_c$, and such that $\tilde{V}(x)$ has appropriately low values for $x \in \Omega_h \cap \Omega_c$. We may choose, for example, the value of $V(x)$ in a nearby minimum or in the global one. In the classically inaccessible parts of $\Omega_h$ we set $\tilde{V}(x) = \infty$, $i.e.$ we allow no nodes there. The nodes are then chosen similiarly to \cite{GarLight}, but using $\tilde{V}(x)$ instead of $V(x)$. Figure \ref{fig1} illustrates $\tilde{V}(x)$ for a Morse potential. In 1-D and 2-D problems this approach should not pose serious difficulty, the ``hard'' potential regions and nearby minima may be 
identified by inspection of the potential surface. In higher dimensions potential surface geometry is not so easy to visualize, yet simple ways may still exist for modifying the choice of
nodes and width parameters. We defer their discussion to section \ref{Conclude}.

\item In \cite{GarLight}, where Gaussian RBFs are used in Galerkin
eigenvalue calculations, it is observed that accuracy improves as the global
width parameter $c$ is decreased until unstable calculations appear. Therefore $cond(S)$ is used as an indicator for a good choice of $c$;
$c$ is chosen so that $cond(S)$ is in the range $10^9-10^{12}$.                
However, it is desirable that our choice of $c$ will relate to the accuracy of our results in a more direct way, and will give a narrower range of possible $c$ values.
We therefore suggest a different approach for choosing $c$.
Let $U$ be a matrix whose columns give (a few or many) approximate eigenfunctions calculated using collocation in
the Gaussian RBF basis. Since the exact eigenfunctions are orthonormal
we suggest to use the ``collocation orthogonalization error'' $\| U^\ast S U - I \|$ as a performance indicator for the choice of $c$ in the collocation approach. To our
surprise the values of $c$ minimizing the collocation orthogonalization error were close
to those minimizing eigenfunction and eigenvalue errors also in the Galerkin approach. Another criterion for choosing $c$ based on Hamiltonian trace minimization is suggested in section \ref{Conclude}.


\item The simplicity of RBF collocation is very appealing. But how does collocation compare to the traditional Galerkin approach? It is reported in \cite{YangPeet1}, \cite{YangPeet2} that for their basis functions collocation accuracy is similiar to that of Galerkin. They therefore conclude that collocation accuracy is similiar to that of Galerkin. However, in all our numerical examples the best Galerkin results were better than the best collocation results.

Another question is how collocation works together with the quasi randomly distributed Gaussians of \cite{GarLight}, $i.e.$ how does performance compare to Gaussians whose centers are
distributed deterministically with the same density?
To the best of our knowledge this was not previously examined in the literature.
In our numerical examples collocation performed generally worse with the quasi random centers.


\end{enumerate}

\section{Numerical examples and discussion}
\label{NumEx}
\begin{figure}
\centering
\includegraphics[width = 3.5in]{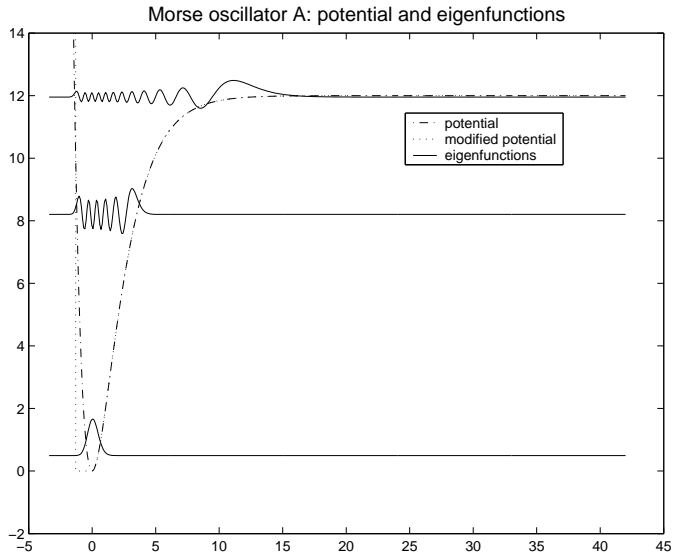}
\caption{The Morse A potential $V(x) = 12 (1 - \eexp^{- 0.5 x})^2$ is
shown here
together with the $0$th, $10$th, and $23$rd eigenfunctions. The
graph of each eigenfunction is translated vertically by its own eigenvalue. The dotted line is the graph of the corresponding modified potential $\tilde{V}(x)$
defined in equation (\ref{ModPot}). 
\label{fig1}}
\end{figure}

The suggestions and observations above were tested on two Hamiltonians which we call
Morse A, and Morse B. Morse A, from \cite{YangPeet1}, has
potential $V(x) = 12 (1 - \eexp^{-0.5 x})^2$ which is plotted in figure \ref{fig1}, and parameters $m = 6$, $\hbar = 1$. This system has $24$ bound states. Morse B
has the potential $V(x) = 12 (1 - \eexp^{-0.9566 x})^2$, and all other parameter values identical to those of Morse A. This system has 13 bound states. Analytic formulas for eigenvalues and eigenfunctions were taken from \cite{MorseHamAnalytic}. 

\begin{figure}
\centering
\includegraphics[width = 3.5in]{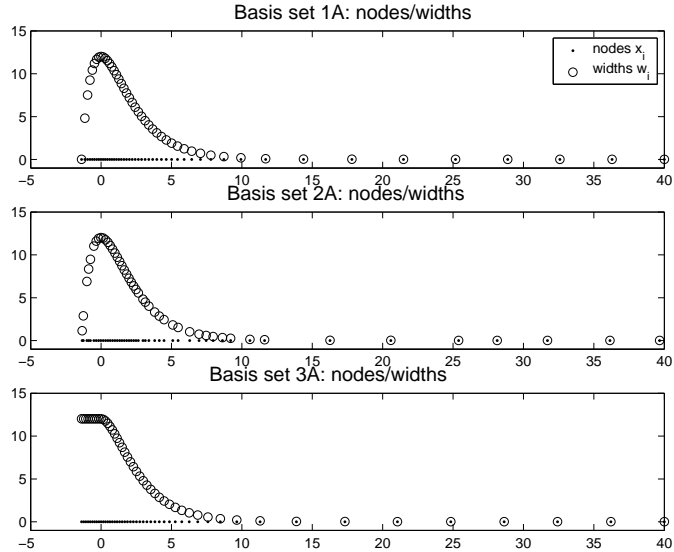}
\caption{Nodes $x_i$ (dots: $\cdot$) and width parameters $w_i$ (circles: $\circ$) used for the different kinds of basis functions. Set 1A: deterministic node choice. Set 2A: quasi random node choice \cite{GarLight}. Set 3A: deterministic node choice with density corresponding to the modified potential in equation (\ref{ModPot}). See section \ref{NumEx} for a full explanation of these parameter choices. 
 \label{fig2}}
\end{figure}

\begin{figure}
\centering
\includegraphics[width = 3.5in]{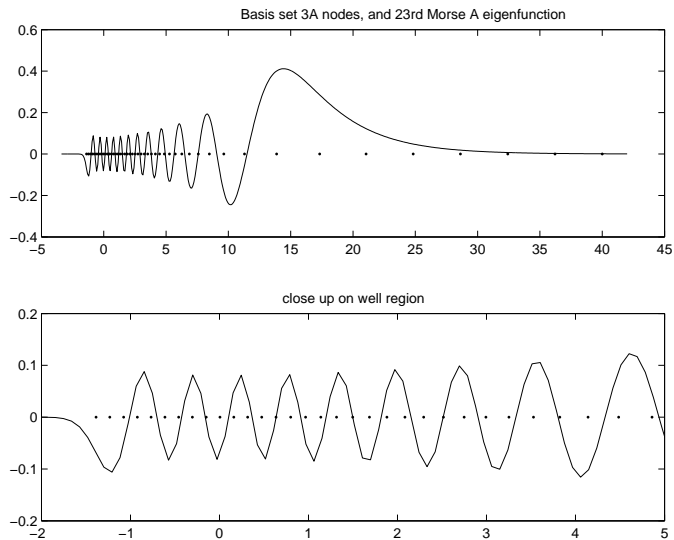}
\caption{Nodes of basis set 3A and $23$rd Morse A eigenfunction. Note the
compatibility between eigenfunction oscillation and node density. The bottom graph
shows a close up on the well region.
 \label{fig3}}
\end{figure}

For each Hamiltonian three sets of Gaussian basis functions were tested. All consisted of $48$
Gaussians, the parameter values $E_{max} = 12$, $\mu = 0.02$ were used
in the construction of all, and all had nodes in an interval $[L,R]$. $L$ is the left classical turning point for $E_{max} = 12$ with values $L \approx -1.3863$, and $L \approx -0.7246$, for system A and B respectively. $R$ was chosen arbitrarily (with a few trials), but well within the region where the highest bound state has decayed close to zero. The values $R = 40$, and $R = 50$, were chosen for system A and B respectively
(in both cases the right classical turning point for $E_{max} = 12$ is too big for our purposes). 
The method for choosing nodes and width parameters in each set of basis functions follows. 
\begin{enumerate}
\item Set 1: Nodes were generated so: $x_1 = R$, $x_{i+1} = x_i - k / \sqrt{E_{max} + \mu - V(x_i)}$, where $k$ was chosen so that $x_{48} = L$. This corresponds to the node density of \cite{GarLight} with $\gamma = 1/2$, see equation (\ref{DensityGL}). Having the nodes, the width parameters were chosen as specified in equation (\ref{WidthsGL}). 
\item Set 2: The nodes were generated quasi randomly as specified in \cite{GarLight}
with the same density as in basis 1. 
The Sobol sequence software package by W. Putsch\"ogl \cite{SobolSoft} was used. Having the nodes, the width parameters were chosen as in basis 1. 
\item Set 3: We regard the left half line $x < 0$ as the hard potential region for both
our Morse potentials. Accordingly, set 3 was generated as set 1, but using the following
modified potential function, which is illustrated in figure \ref{fig1}, 
\be \tilde{V}(x) = \left\{ \begin{array}{cl}
	V(x) & \qquad x > 0  \\
	0    & \qquad L \leq x \leq 0 
\end{array}     \right. \ . \label{ModPot} \ee
\end{enumerate}
The nodes and width parameters of sets 1A, 2A, 3A, are illustrated in figure
\ref{fig2} (this obvious notation distinguishes between basis sets for Hamiltonians A and B). The nodes of set 3A are illustrated in figure \ref{fig3}
together with the $23$rd eigenfunction, the compatibility between node 
density and eigenfunction oscillation is evident.
The node density given by equation (\ref{DensityGL}) with $\gamma = 1$ was also tested, but results were better with $\gamma = 1/2$ as above. 

\begin{figure}[p]
\centering
\includegraphics[width = 2.8in]{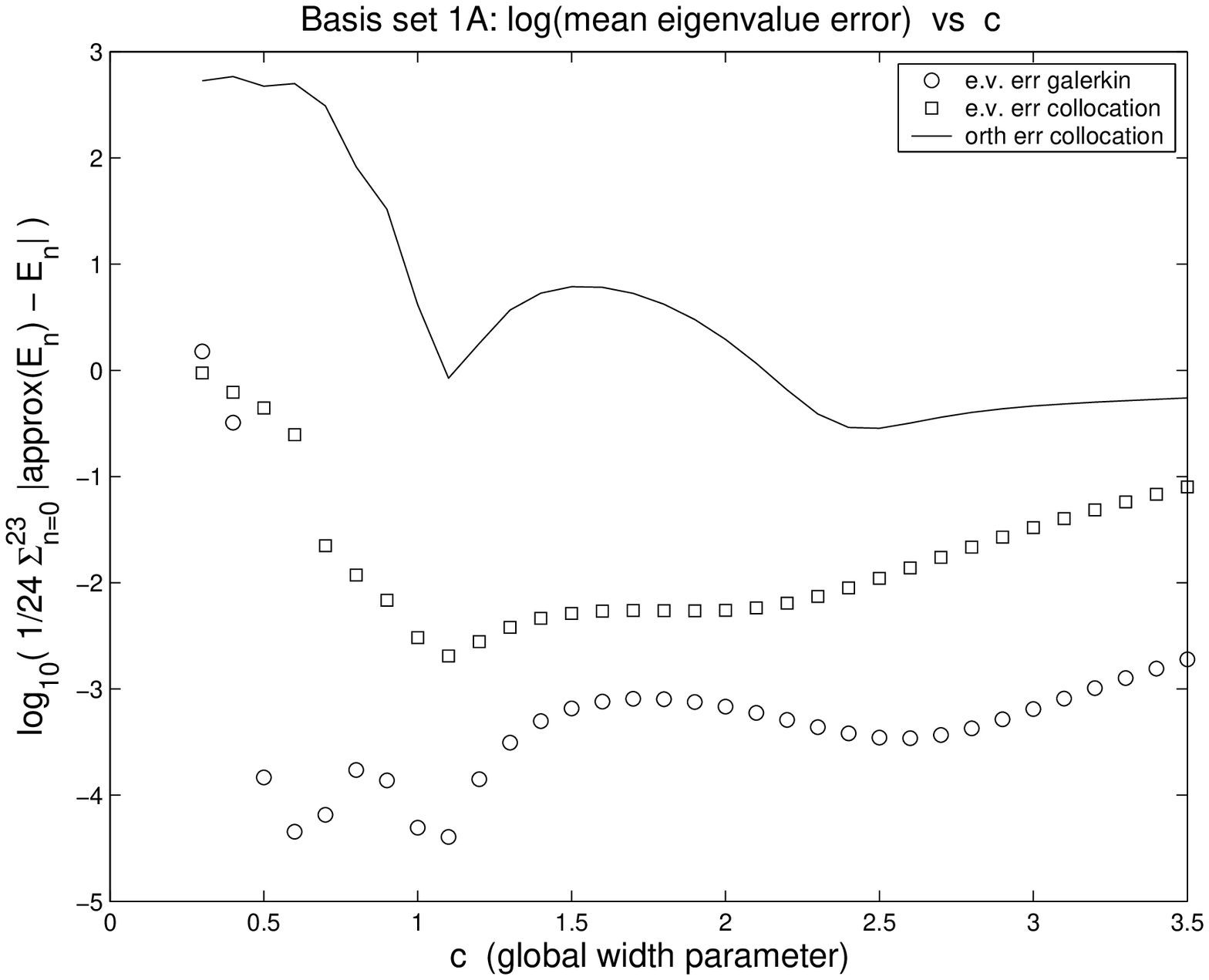}
\includegraphics[width = 2.8in]{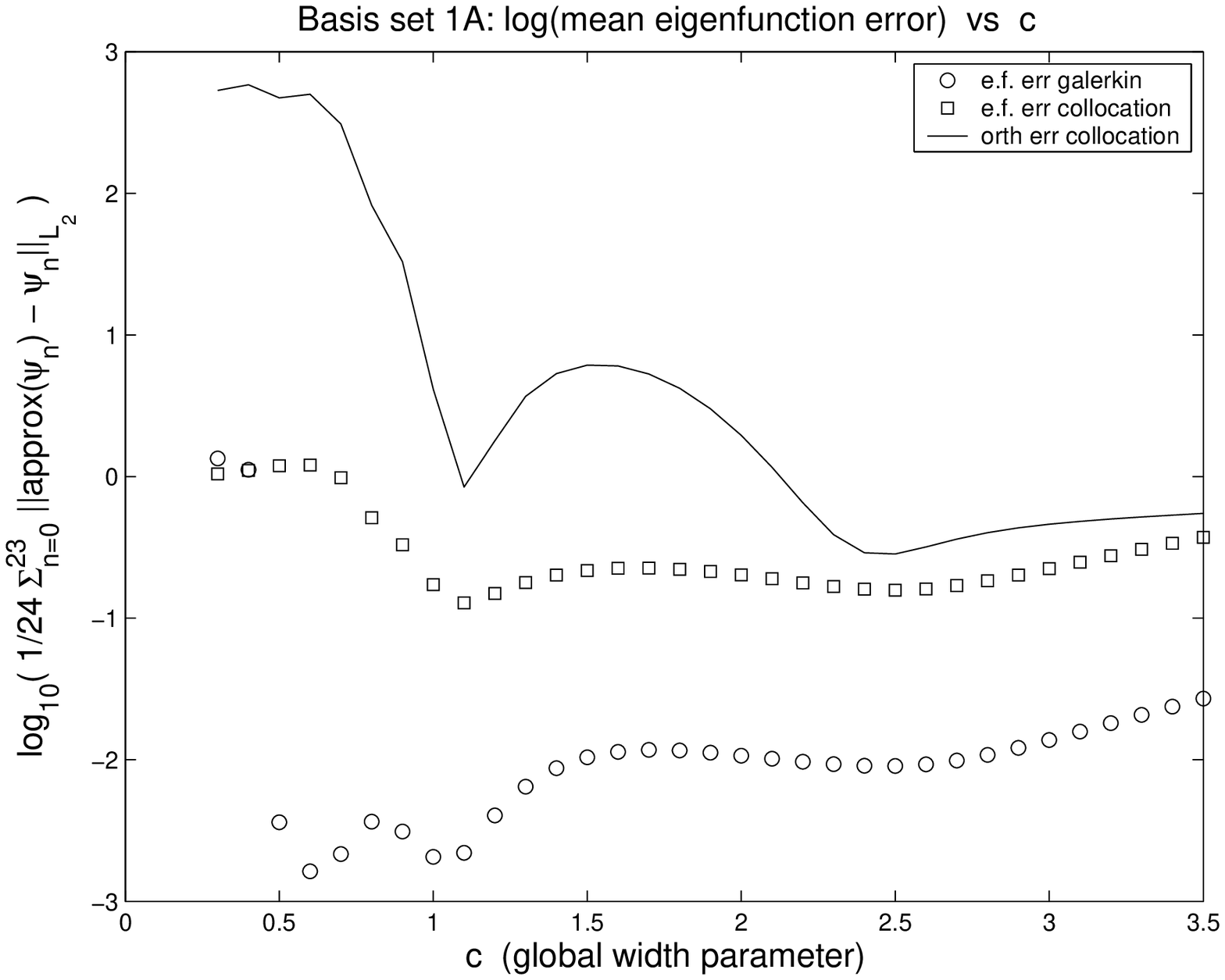}
\includegraphics[width = 2.8in]{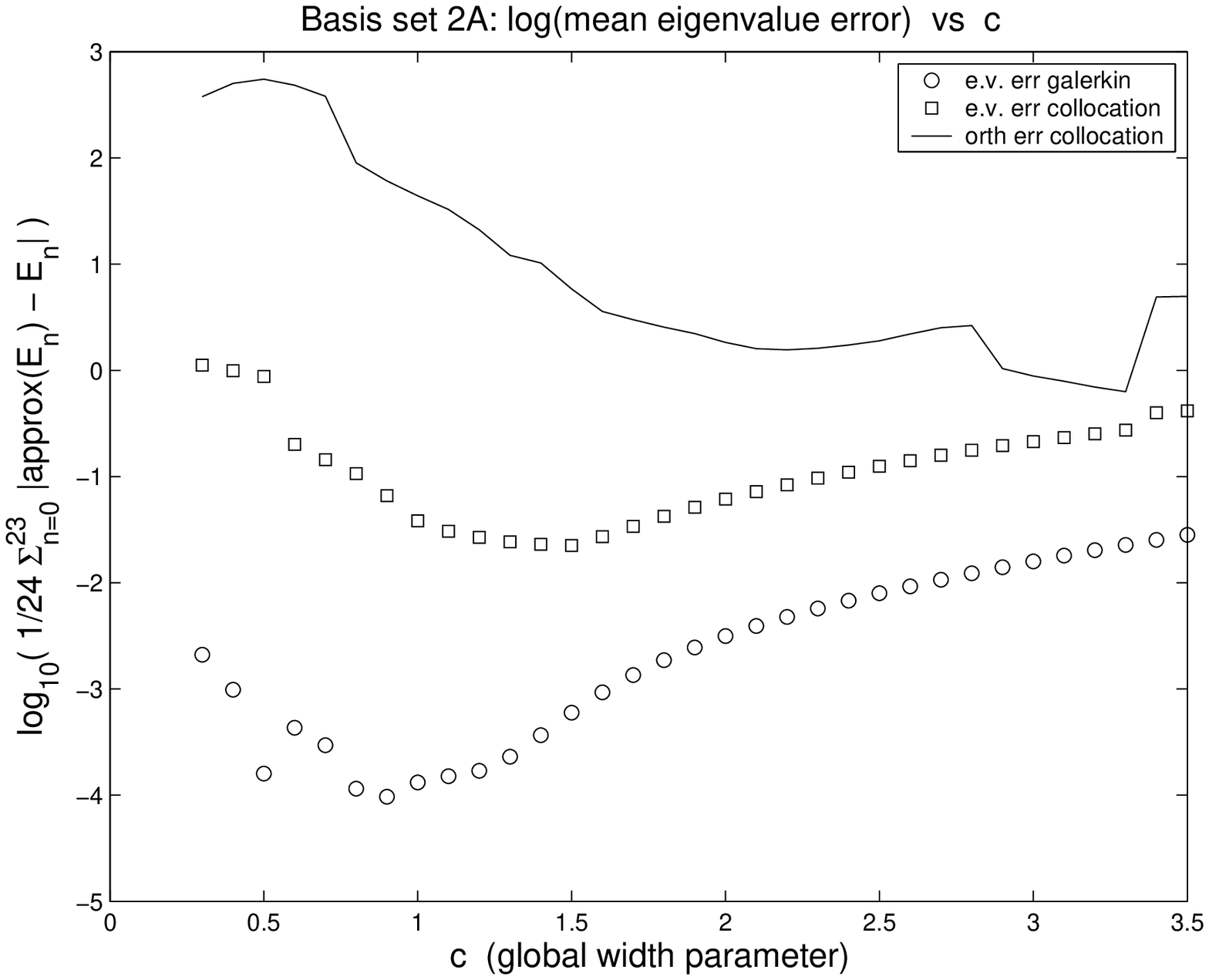}
\includegraphics[width = 2.8in]{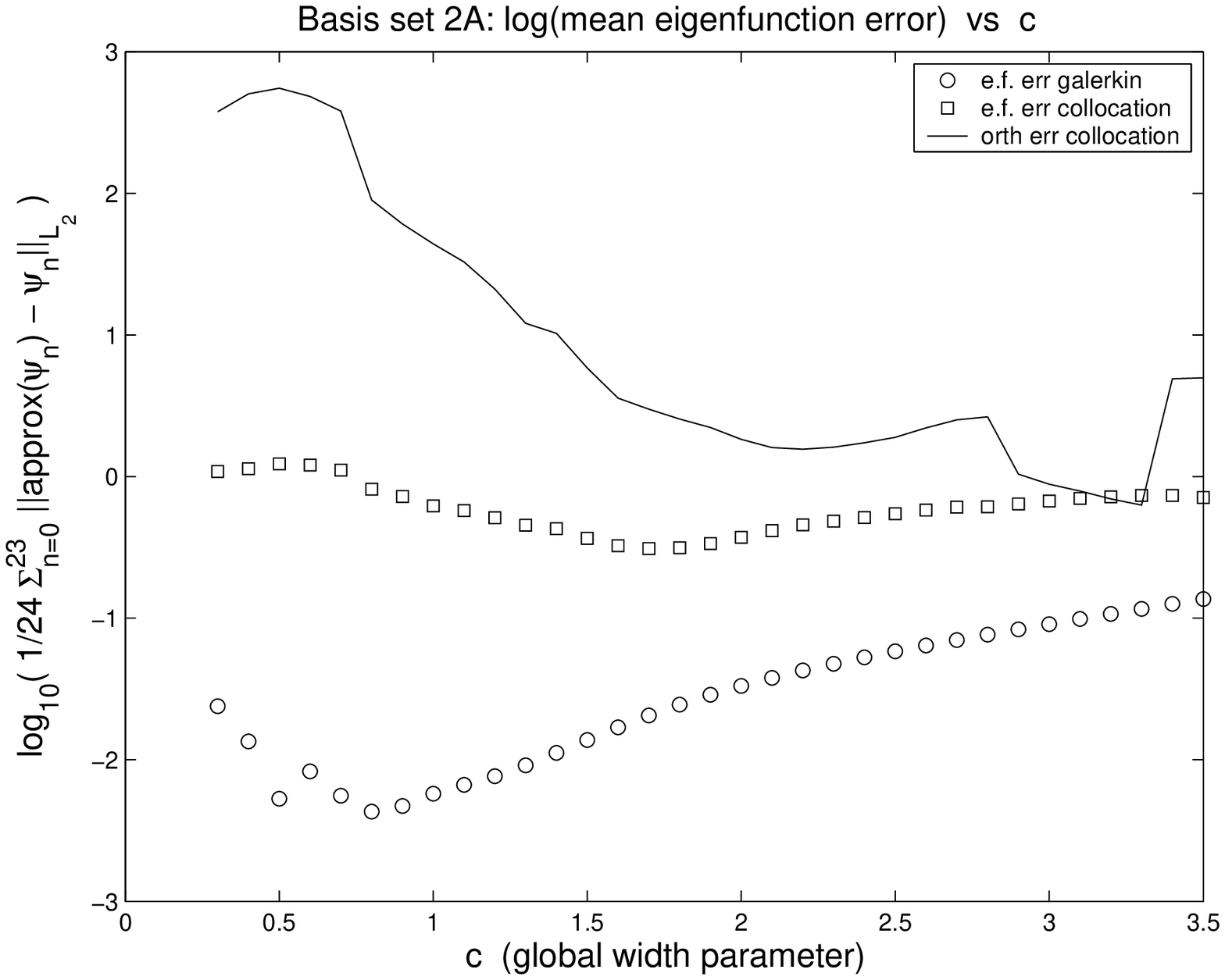}
\includegraphics[width = 2.8in]{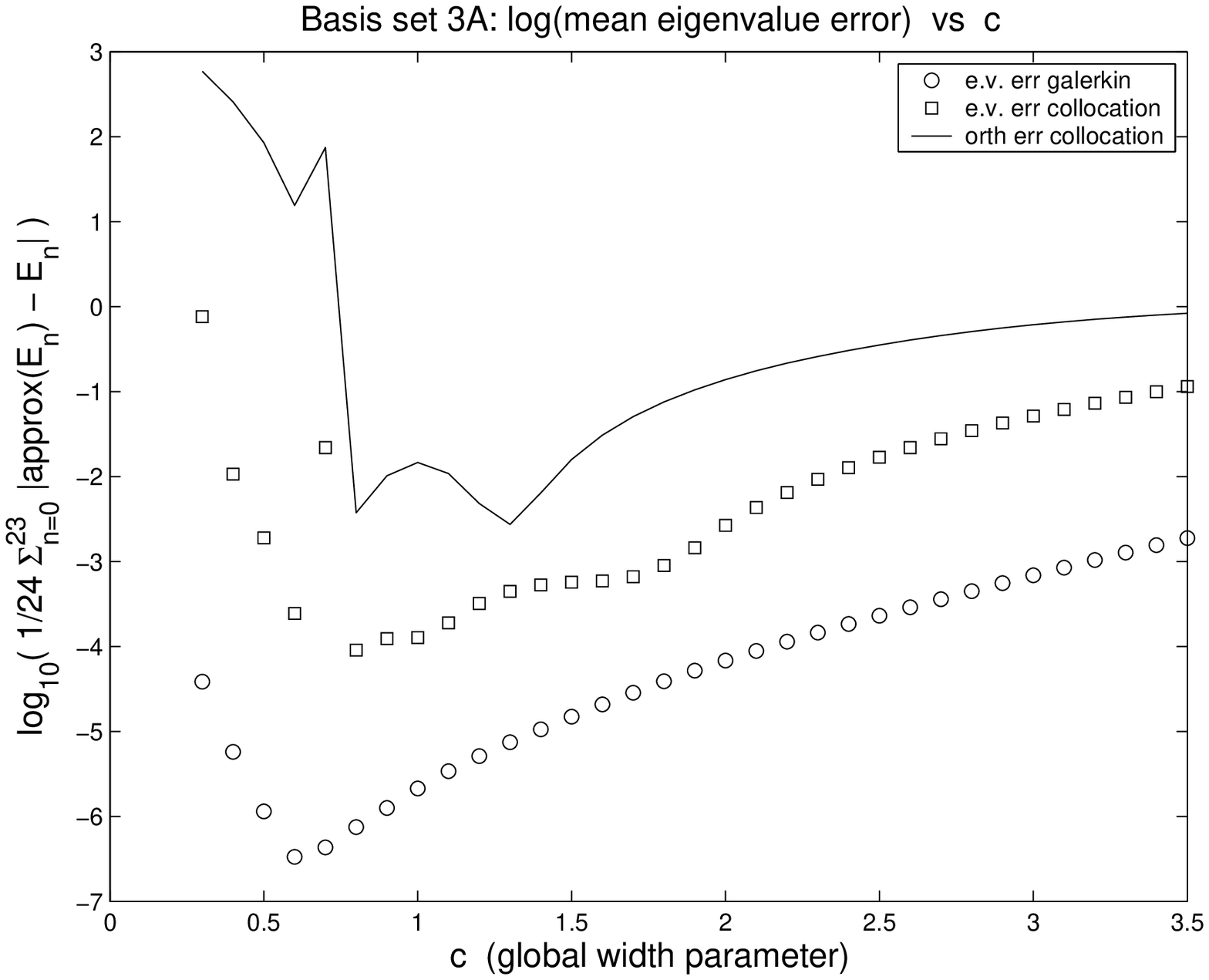}
\includegraphics[width = 2.8in]{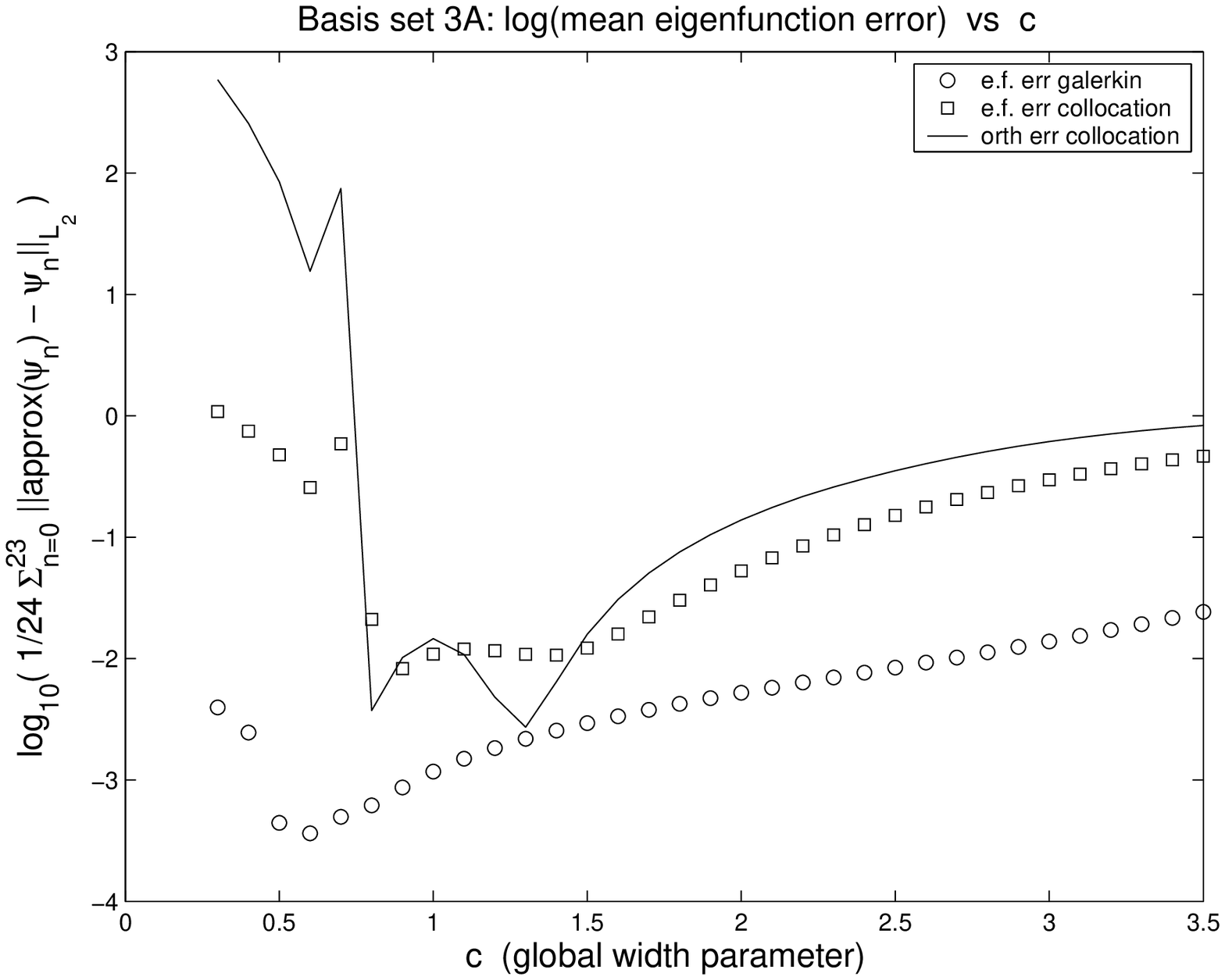}
\caption{Error dependence on global width parameter, Morse A: Logarithms of mean eigenvalue (left column) and eigenfunction
(right column) errors plotted against the global width parameter for the different basis sets (rows), and discretization methods, applied to Morse Hamiltonian A. Circles-Galerkin, squares-collocation,
continuos line-``orth error'' which is defined in equation (\ref{CollOrthErr}). See table \ref{TabOne} A for information on the condition numbers of $S$ and $\Phi$. 
 \label{fig4}}
\end{figure}

\begin{figure}[p]
\centering
\includegraphics[width = 2.75in]{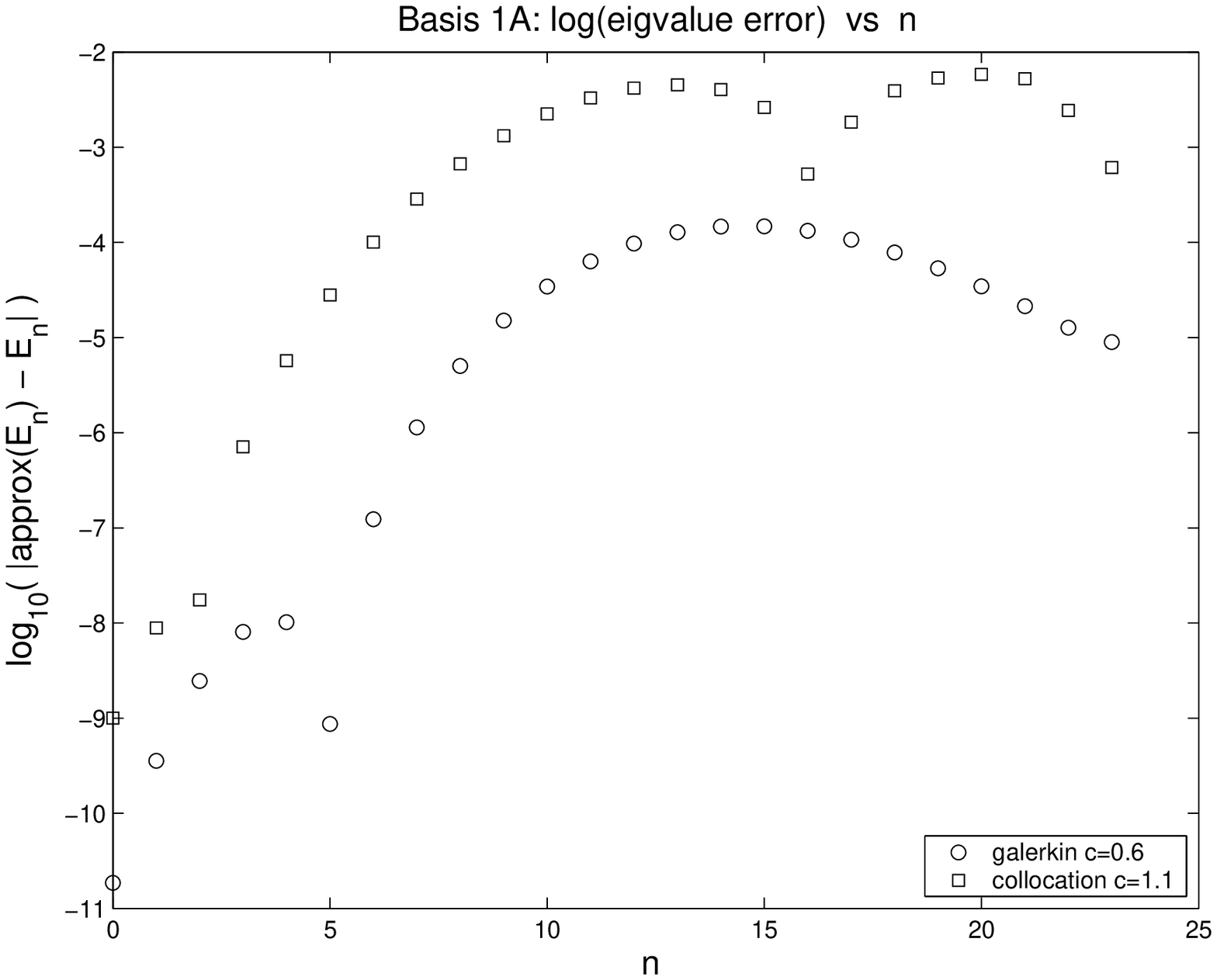}
\includegraphics[width = 2.75in]{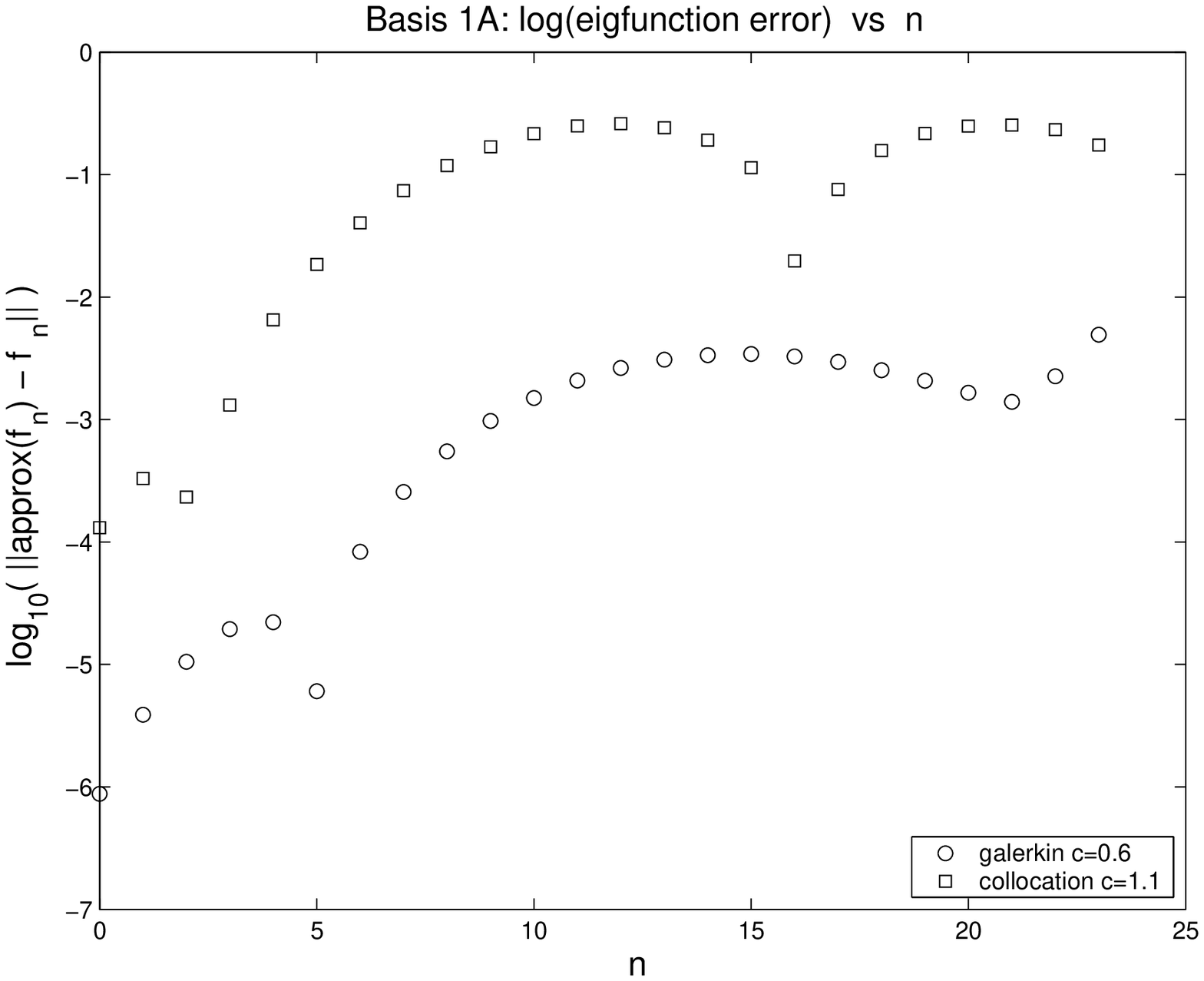}
\includegraphics[width = 2.75in]{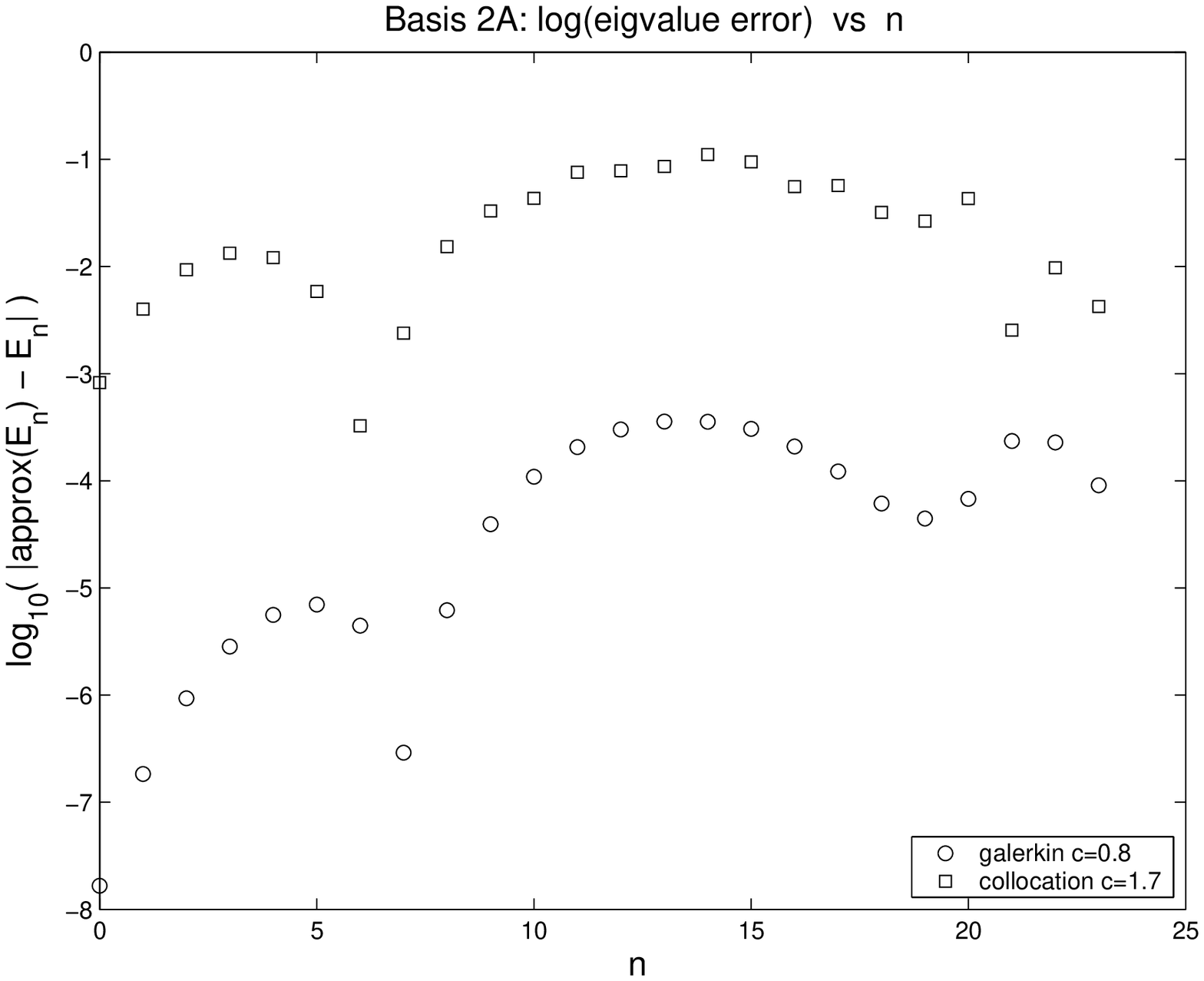}
\includegraphics[width = 2.75in]{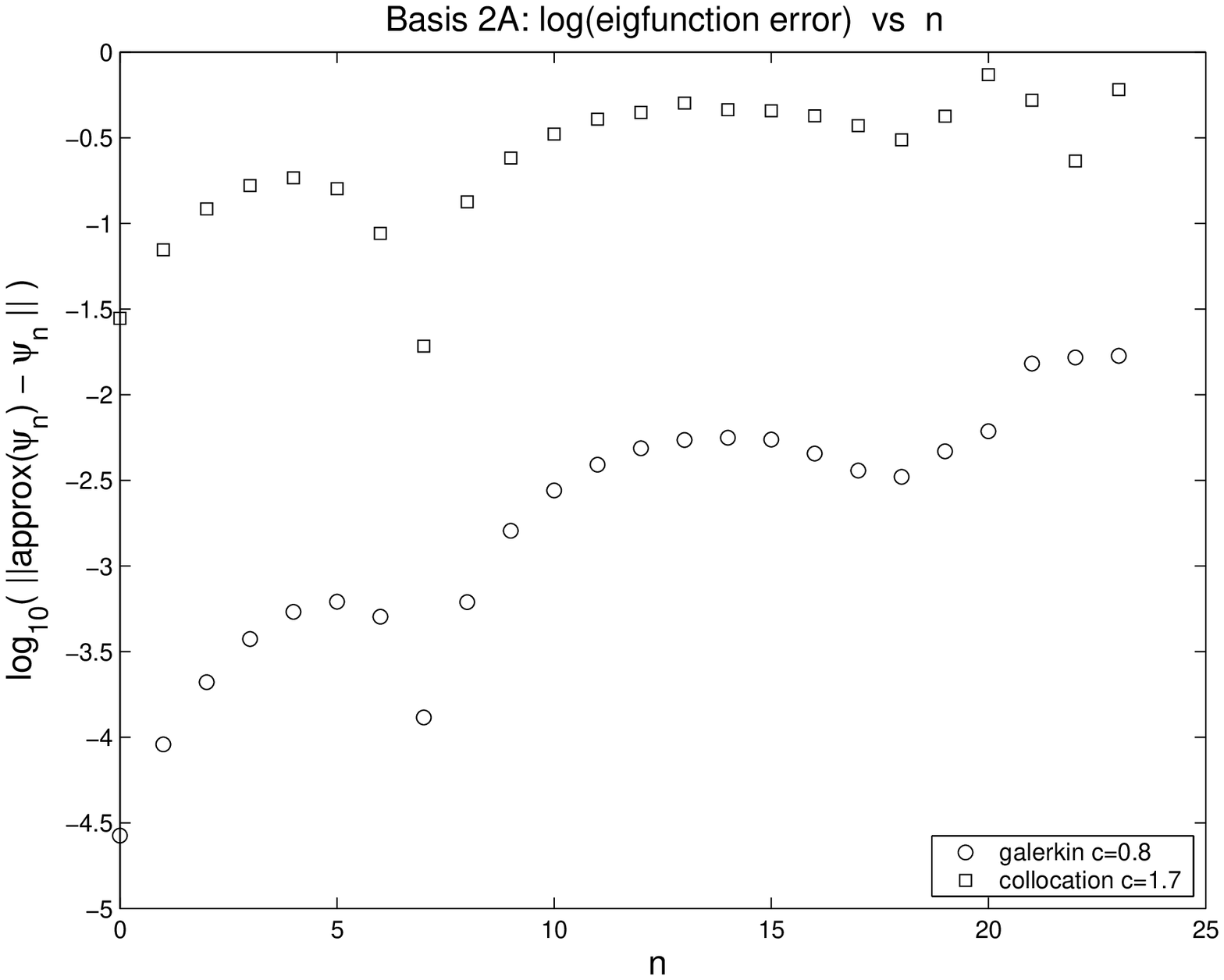}
\includegraphics[width = 2.75in]{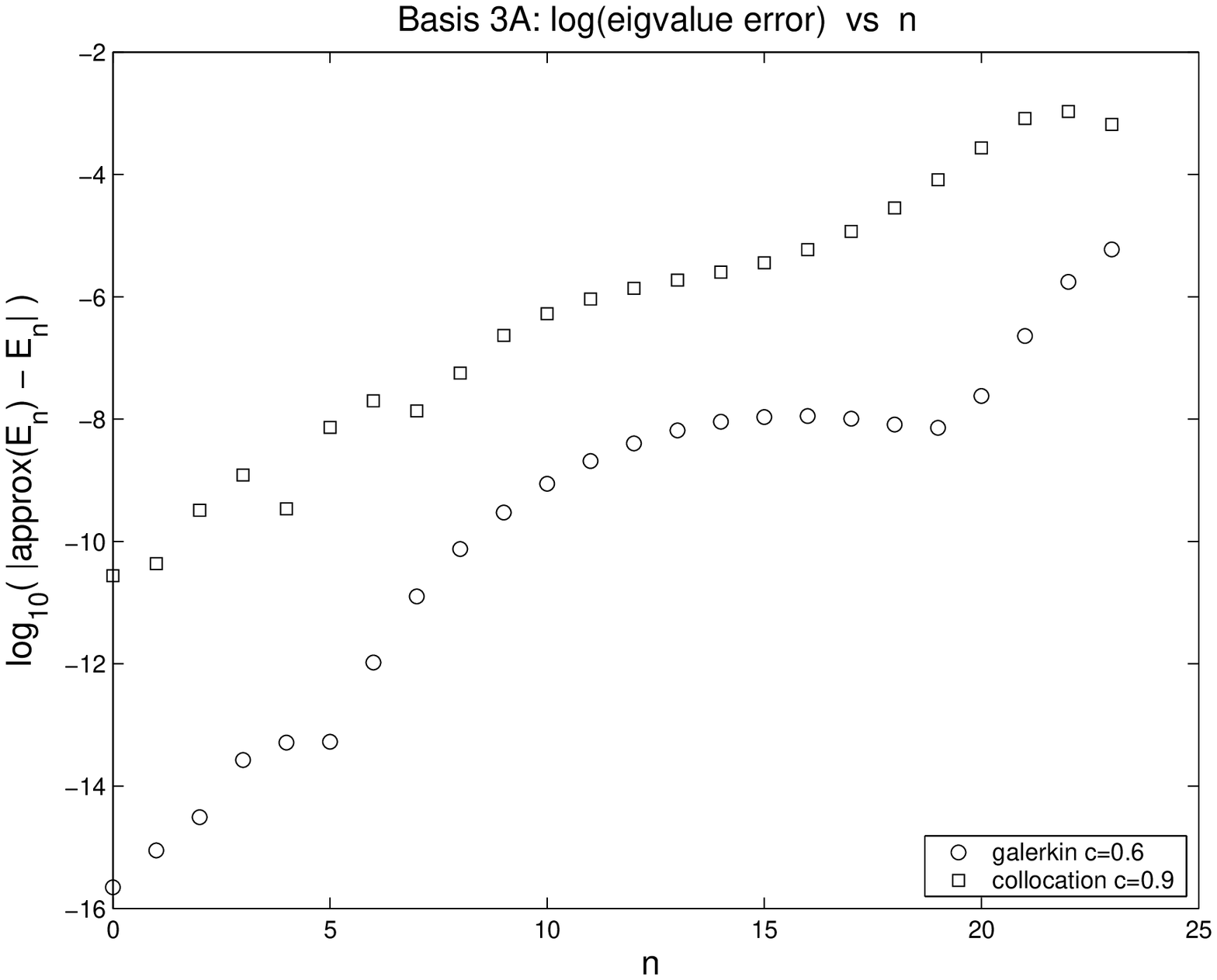}
\includegraphics[width = 2.75in]{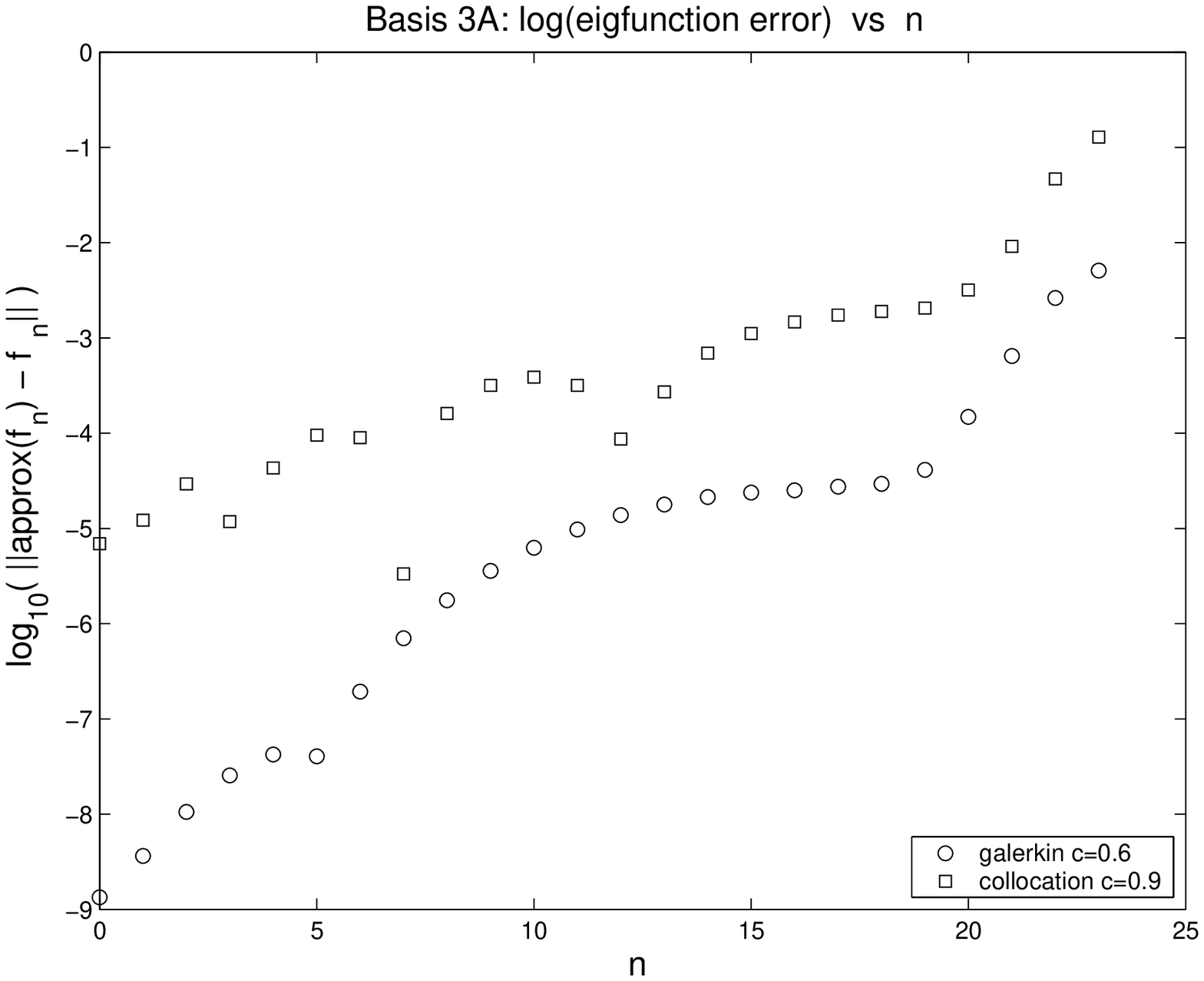}
\caption{Comparing collocation and Galerkin performance, Morse A: Logarithms of eigenvalue (left column) and eigenfunction (right column) errors vs. index for each basis set (rows) and each discretization method
(Circles-Galerkin, squares-collocation) applied to Morse Hamiltonian A. In each case results
were obtained using the global width parameter value which minimizes the mean
eigenfunction error $\epsilon_f$, see table \ref{TabOne} A. 
Note the low errors in the Galerkin/basis set 3 combination. \label{fig5} }
\end{figure}

\begin{figure}[p]
\centering
\includegraphics[width = 2.8in]{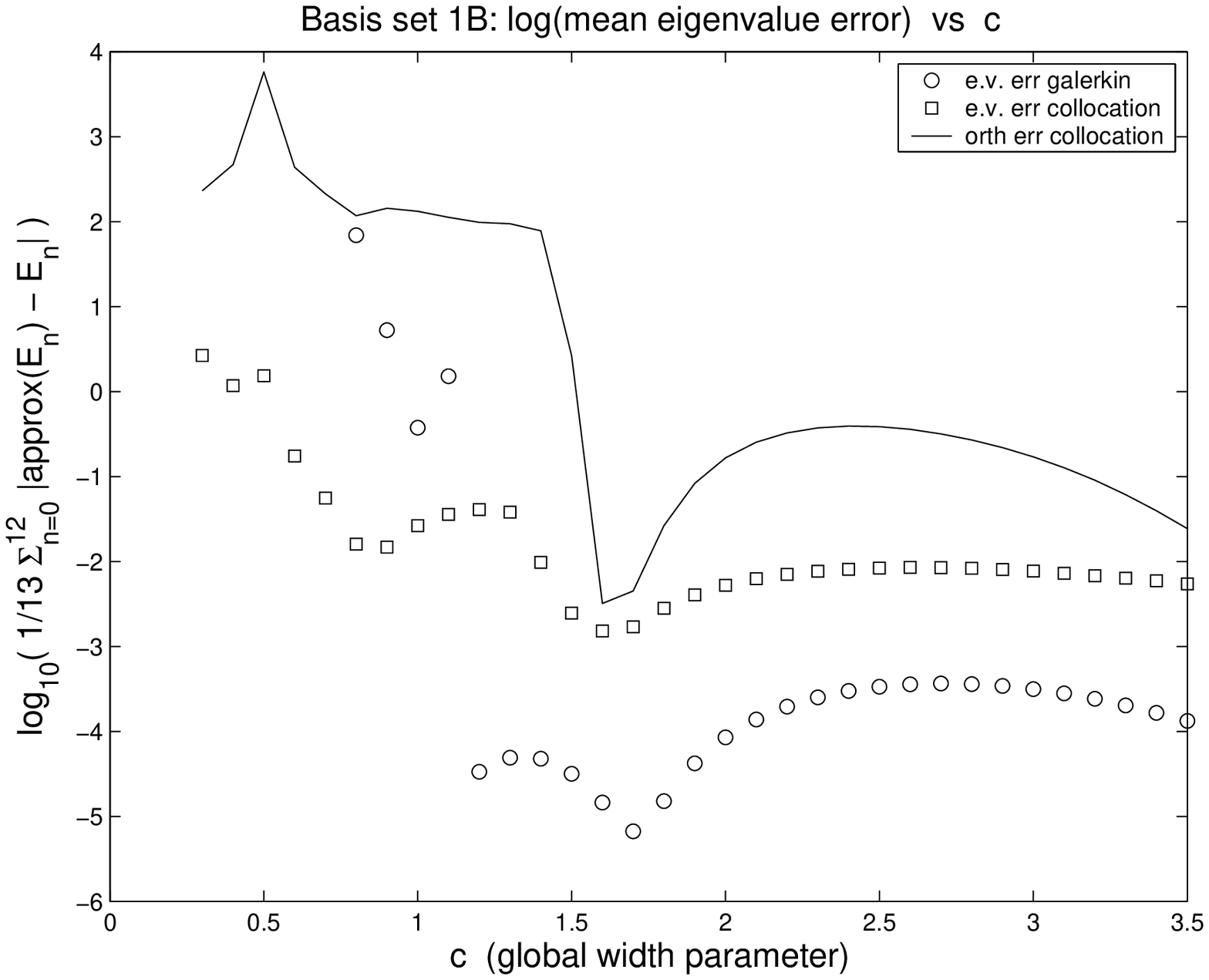}
\includegraphics[width = 2.8in]{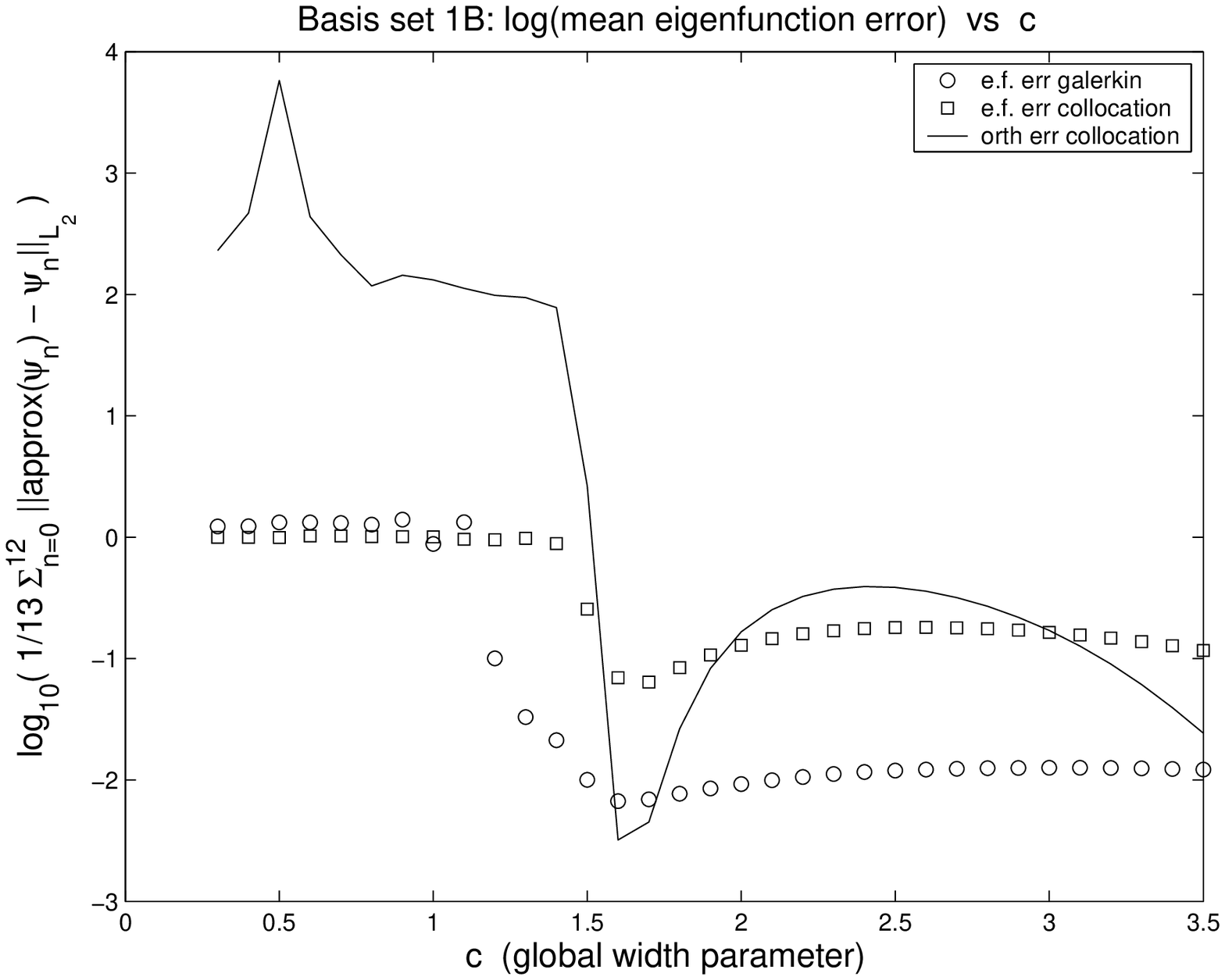}
\includegraphics[width = 2.8in]{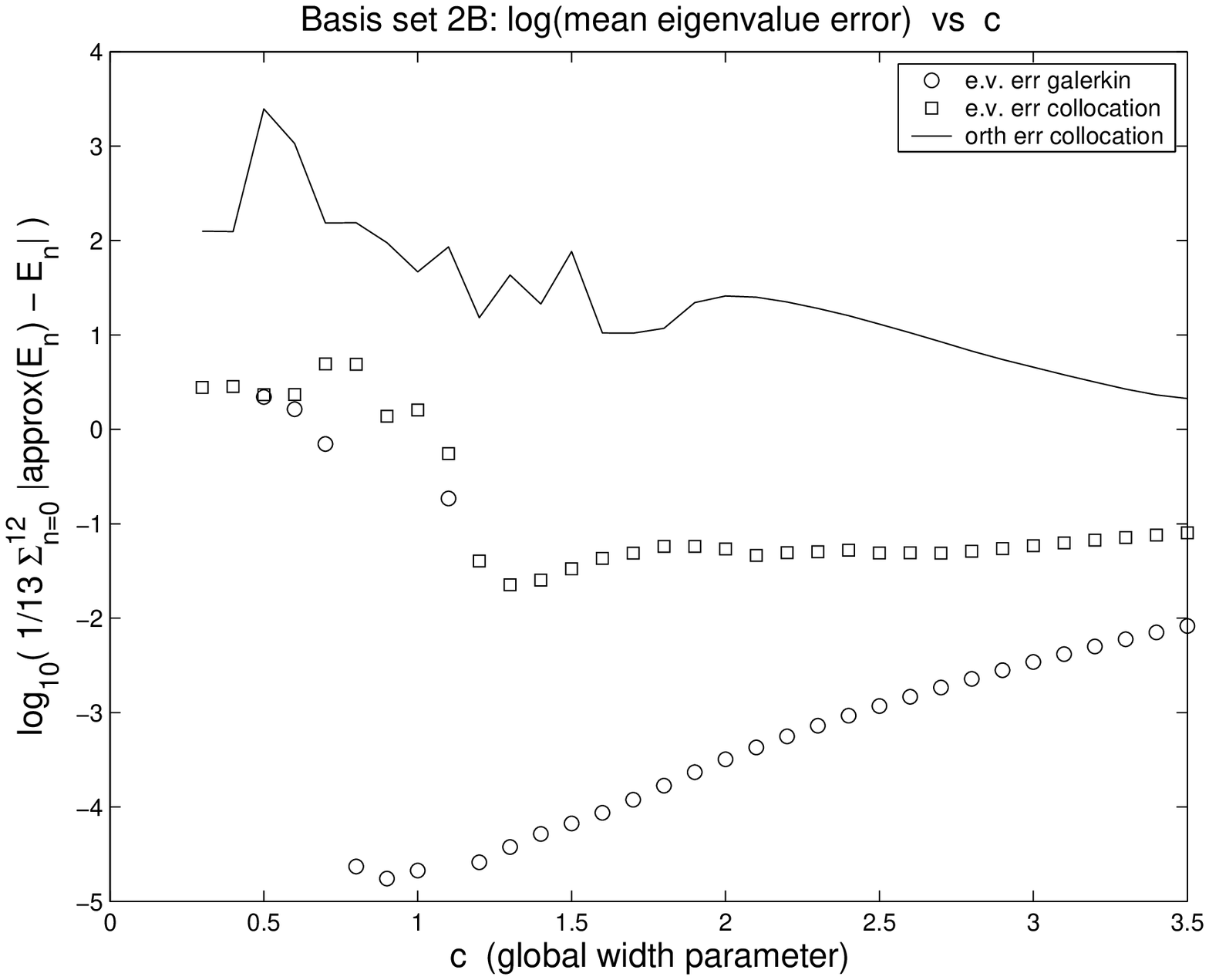}
\includegraphics[width = 2.8in]{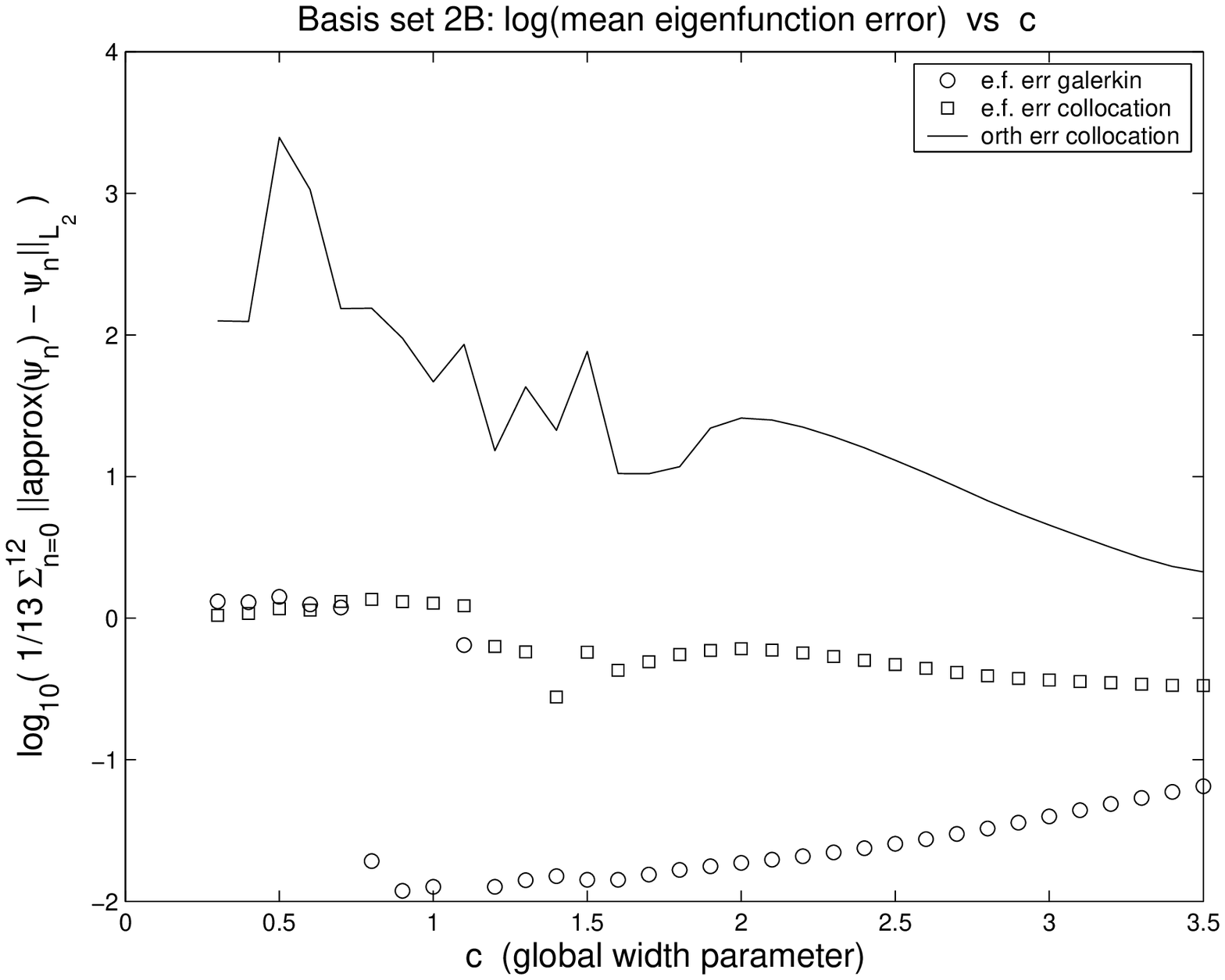}
\includegraphics[width = 2.8in]{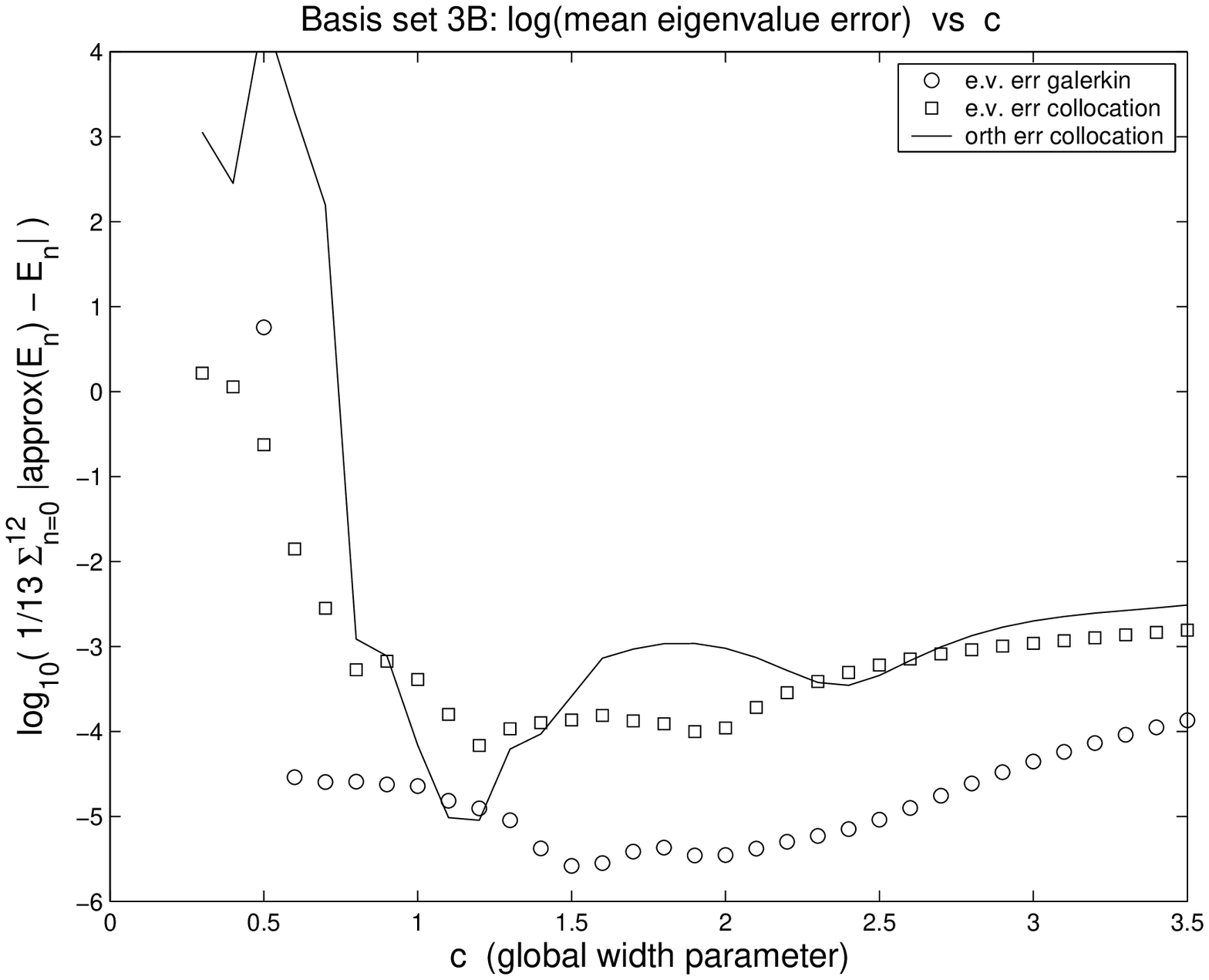}
\includegraphics[width = 2.8in]{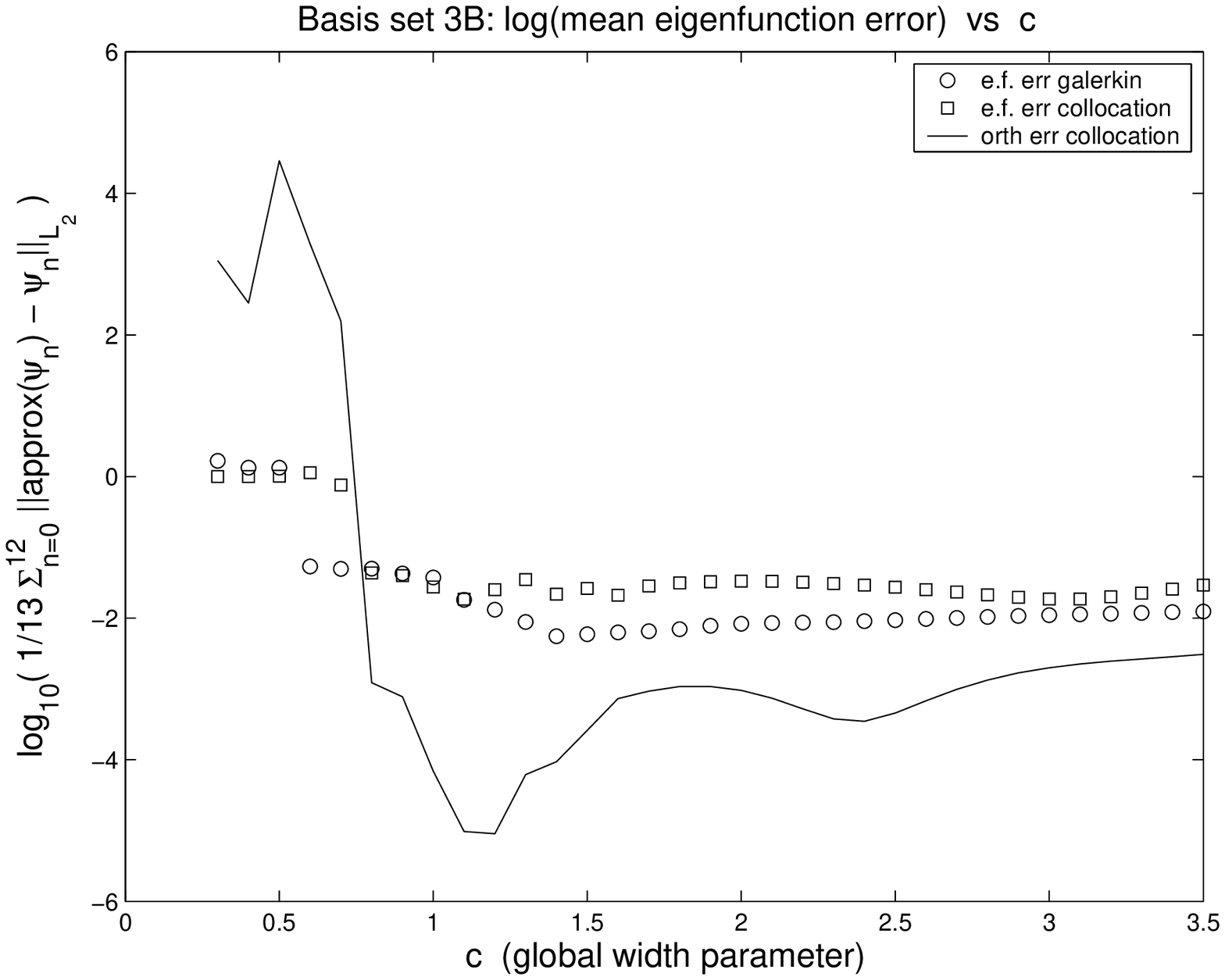}
\caption{ Error dependence on global width parameter, Morse B: Logarithms of mean eigenvalue (left column) and eigenfunction
(right column) errors plotted against the global width parameter for the different basis sets (rows), and discretization methods, applied to Morse Hamiltonian B. Circles-Galerkin, squares-collocation,
continuos line-``orth error'' which is defined in equation (\ref{CollOrthErr}). See table \ref{TabOne} B for information on the condition numbers of $S$ and $\Phi$. 
  \label{fig6} }
\end{figure}

\begin{figure}[p]
\centering
\includegraphics[width = 2.7in]{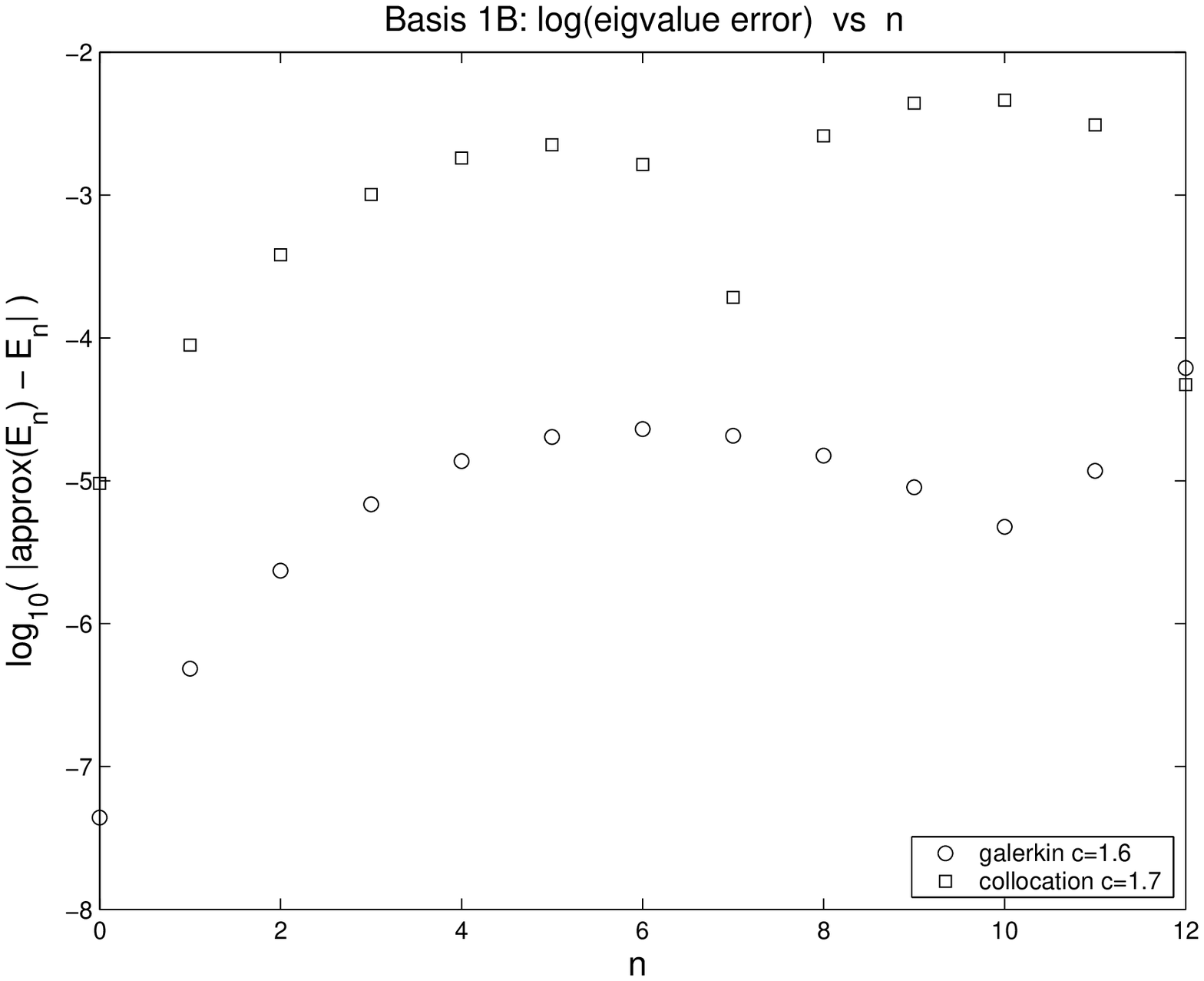}
\includegraphics[width = 2.7in]{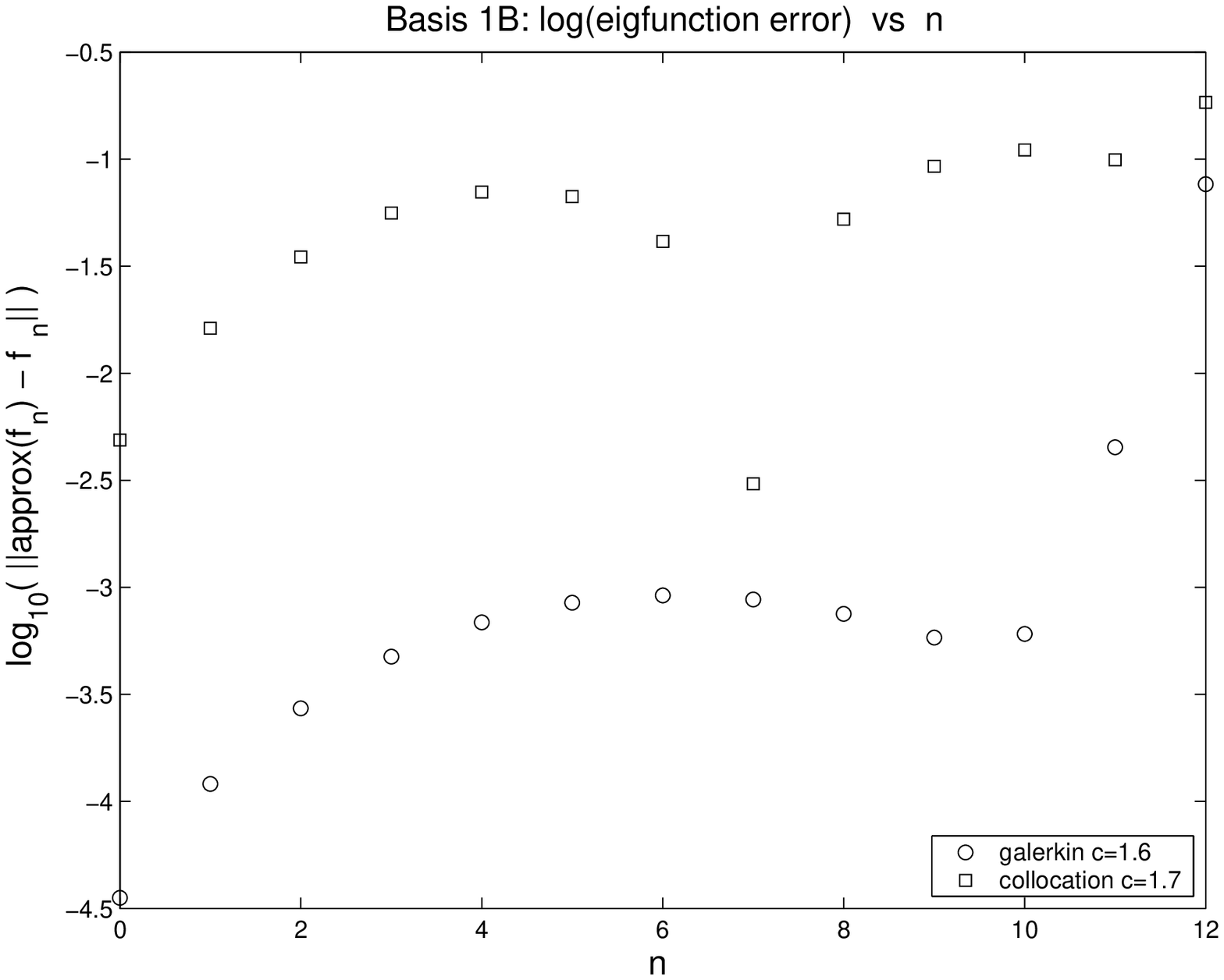}
\includegraphics[width = 2.7in]{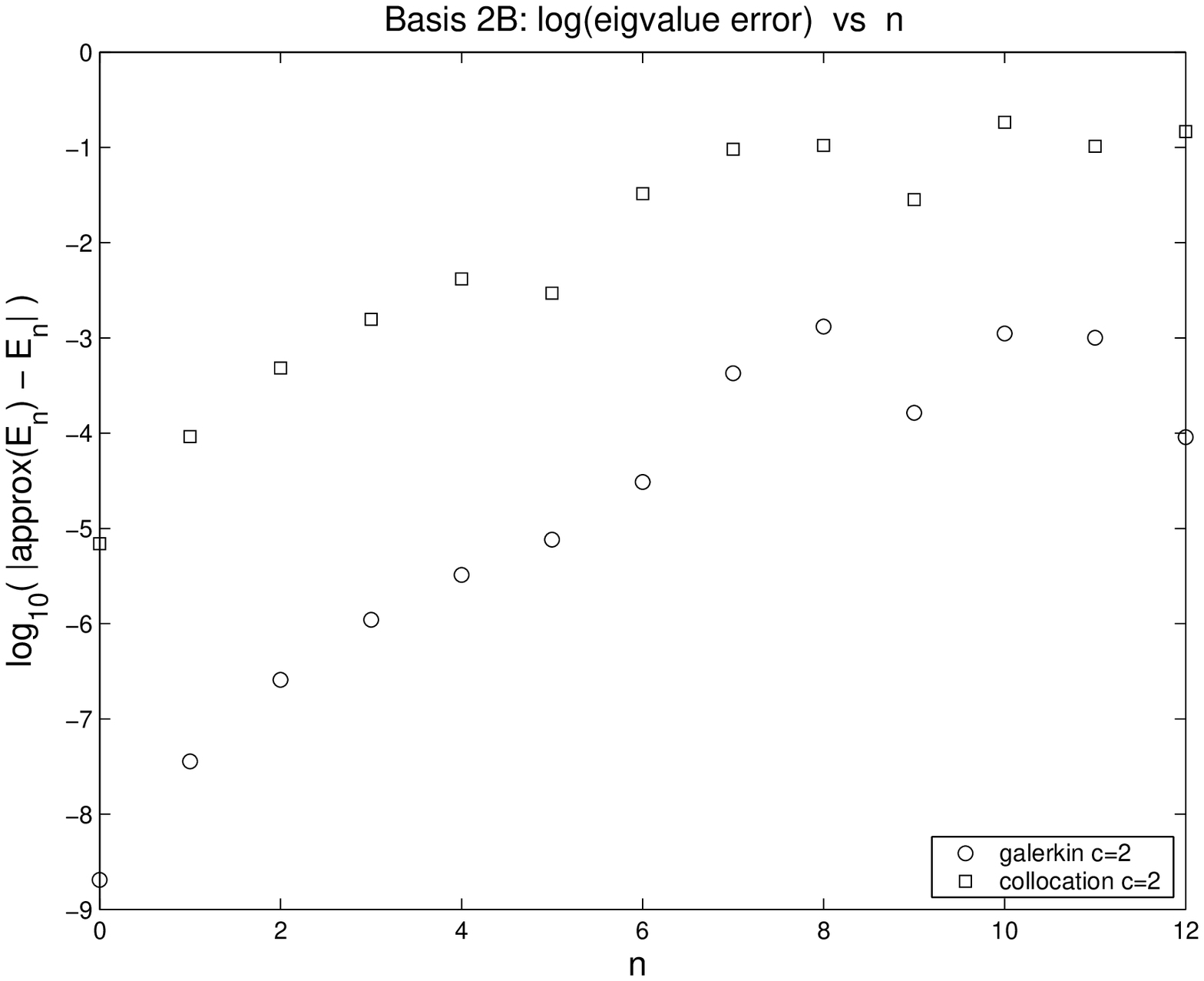}
\includegraphics[width = 2.7in]{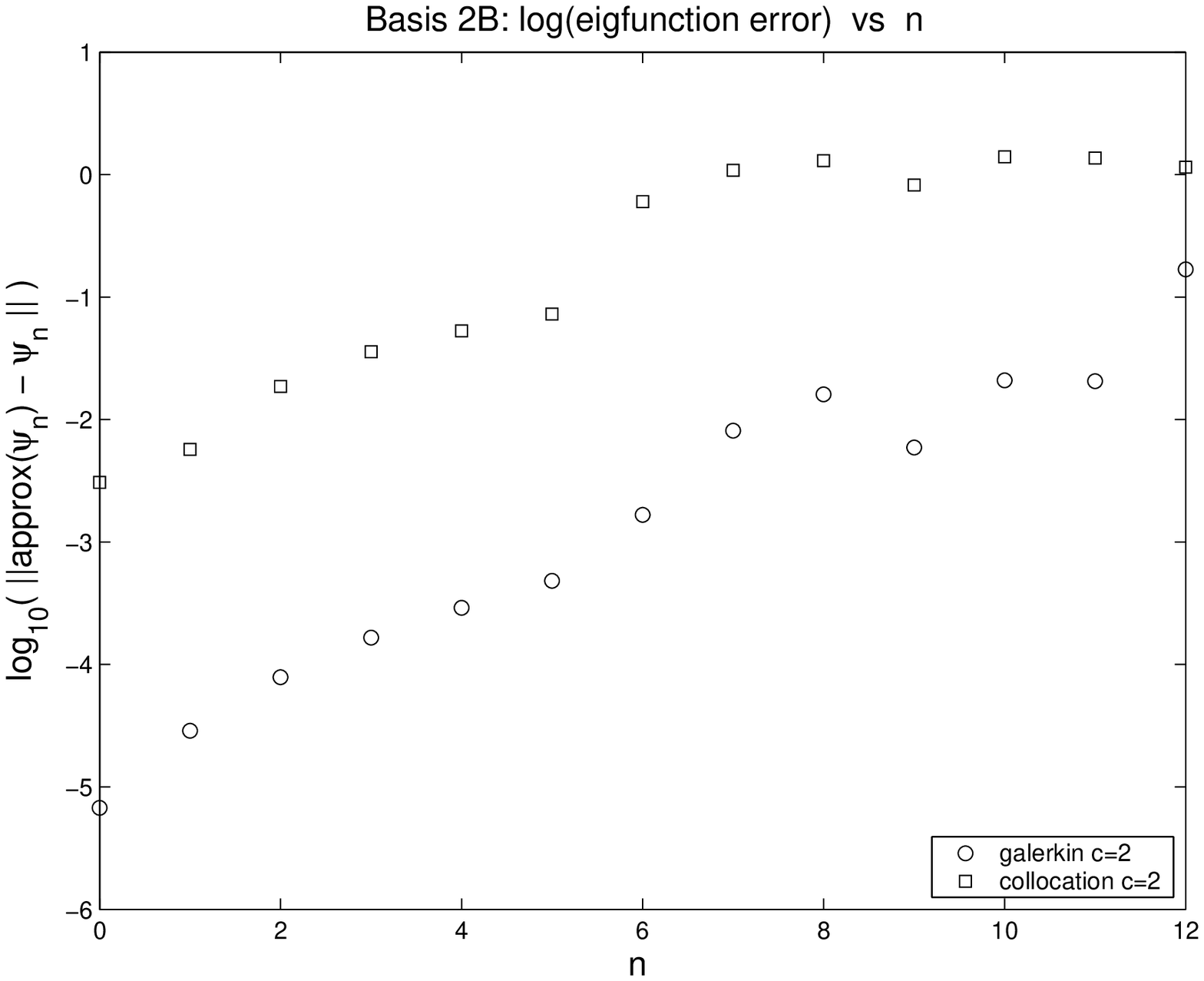}
\includegraphics[width = 2.7in]{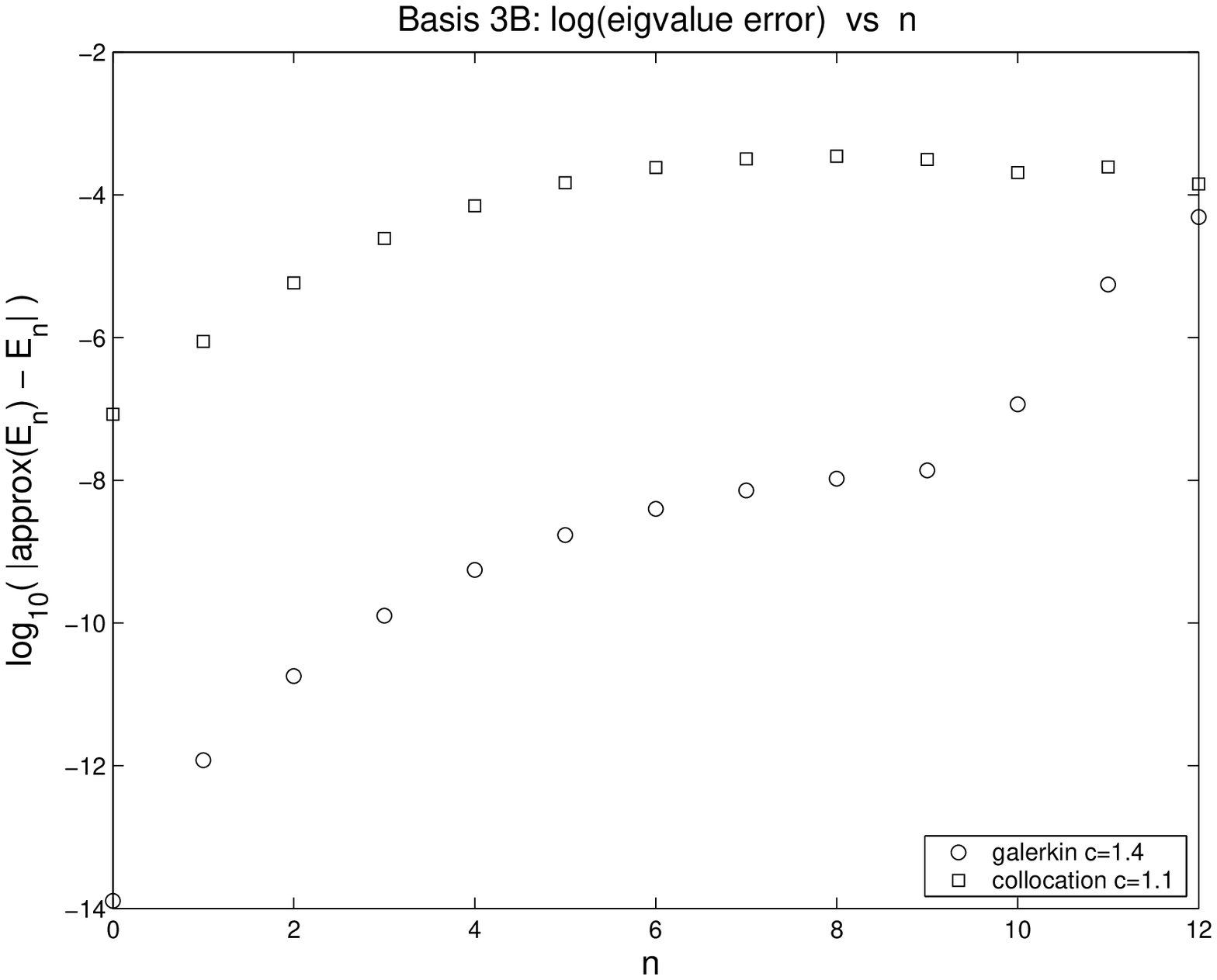}
\includegraphics[width = 2.7in]{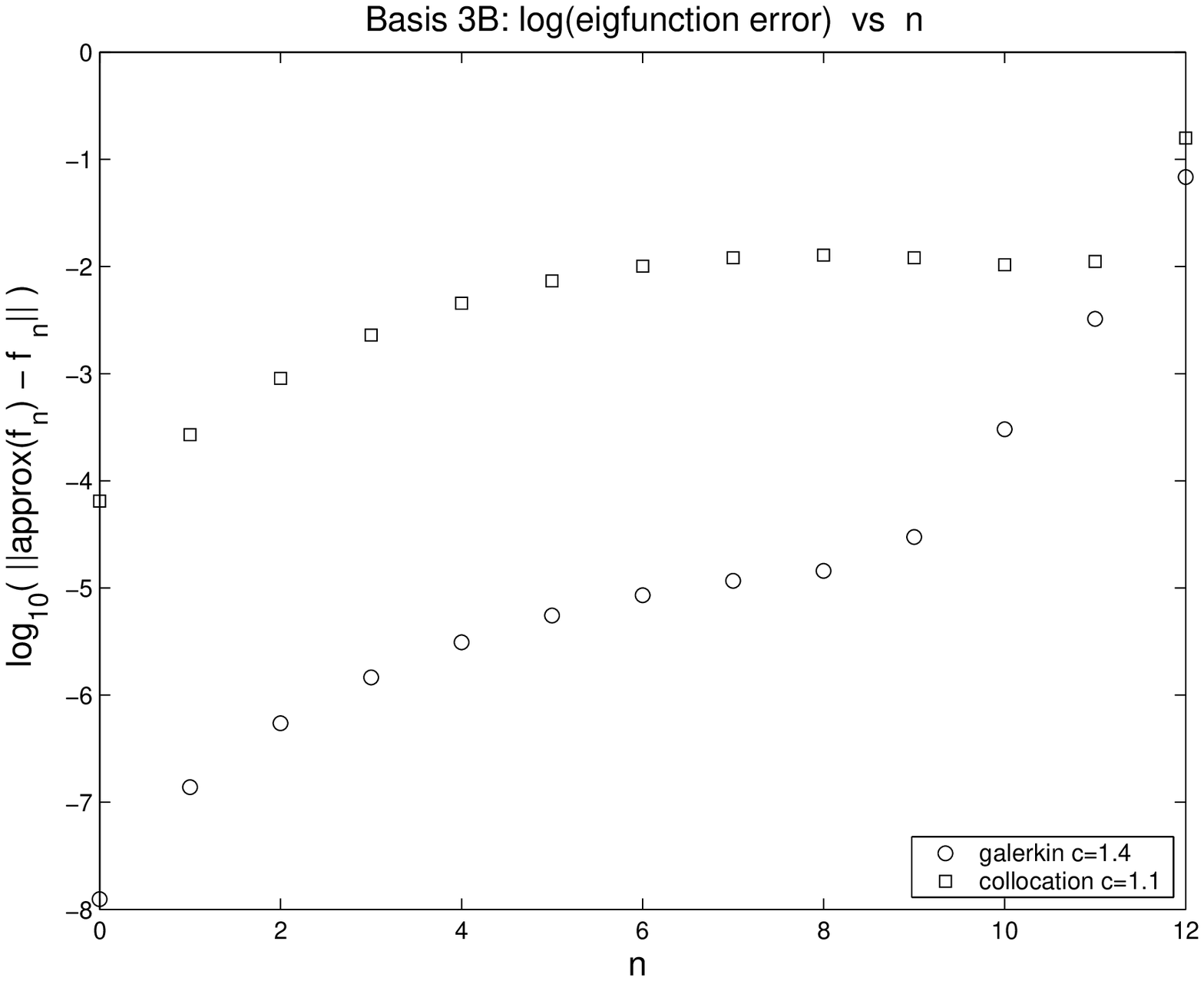}
\caption{Comparing collocation and Galerkin performance, Morse B: Logarithms of eigenvalue (left column) and eigenfunction (right column) errors vs. index for each basis set (rows) and each discretization method
(Circles-Galerkin, squares-collocation) applied to Morse Hamiltonian B. In each case (with the exception of basis 2B) results
were obtained using the global width parameter value which minimizes the mean eigenfunction error $\epsilon_f$, see table \ref{TabOne} B. 
Note the low errors in the Galerkin/basis set 3 combination. 
 \label{fig7}}
\end{figure}

\begin{figure}[p]
\centering
\includegraphics[width = 2.8in]{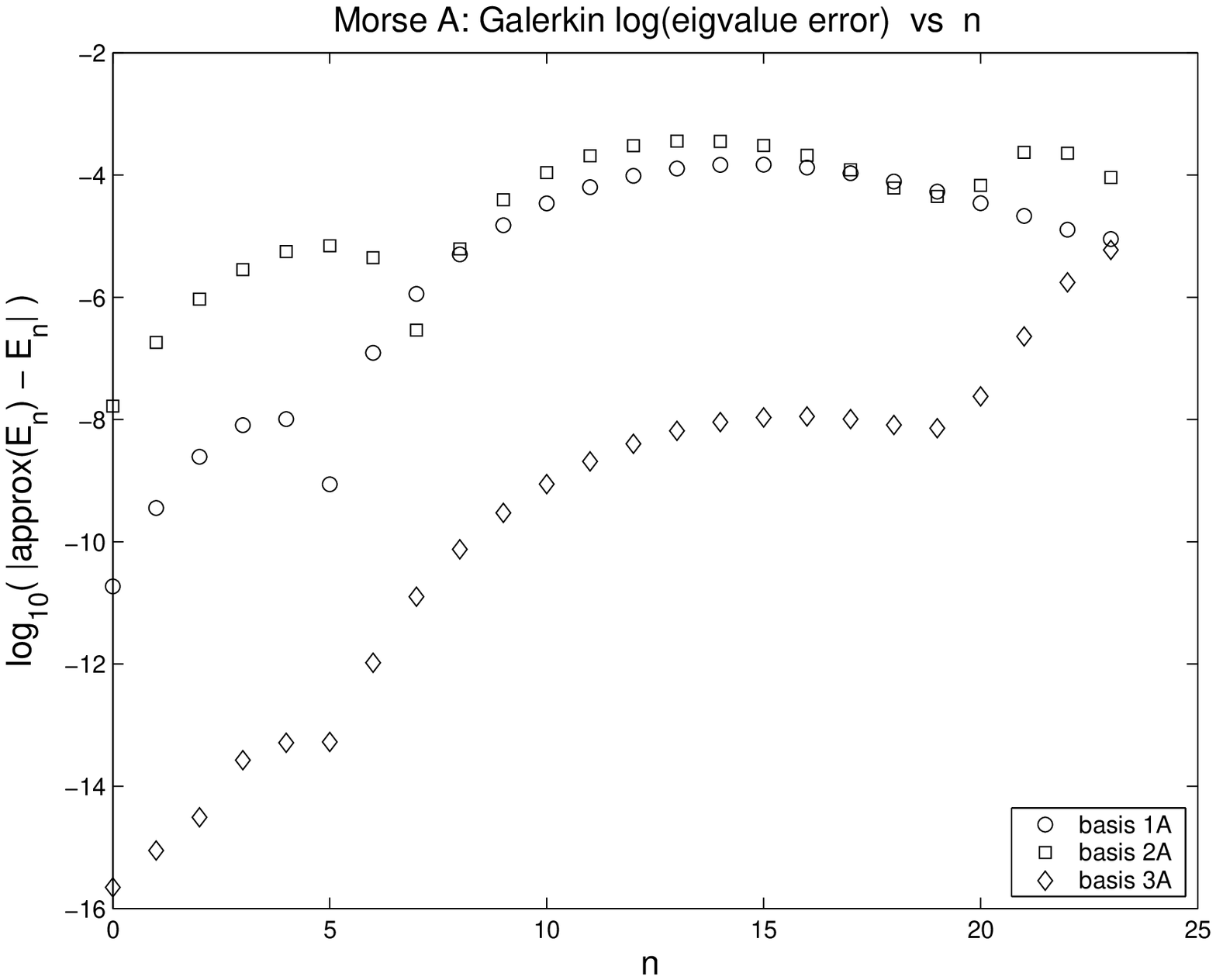}
\includegraphics[width = 2.8in]{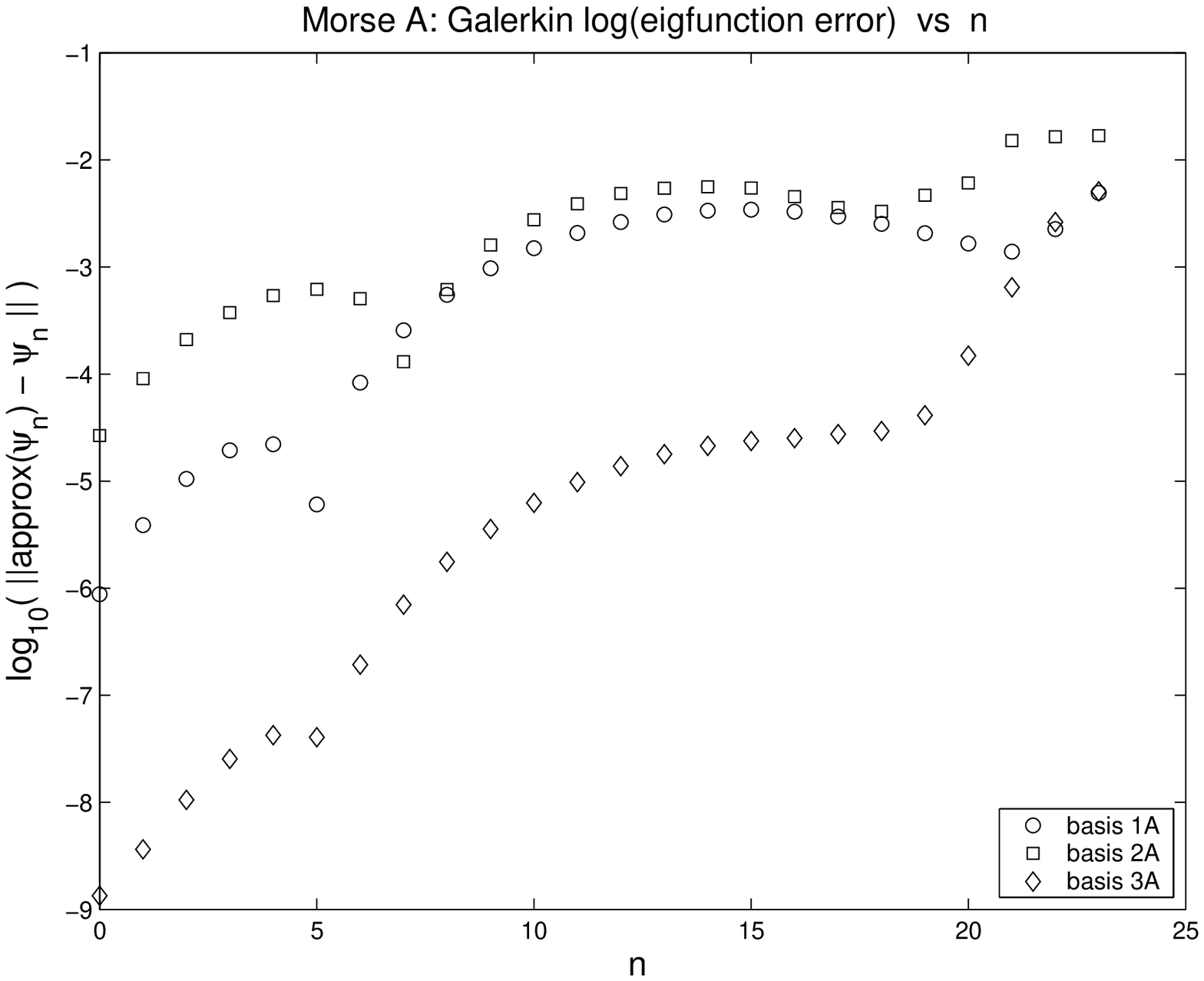}
\includegraphics[width = 2.8in]{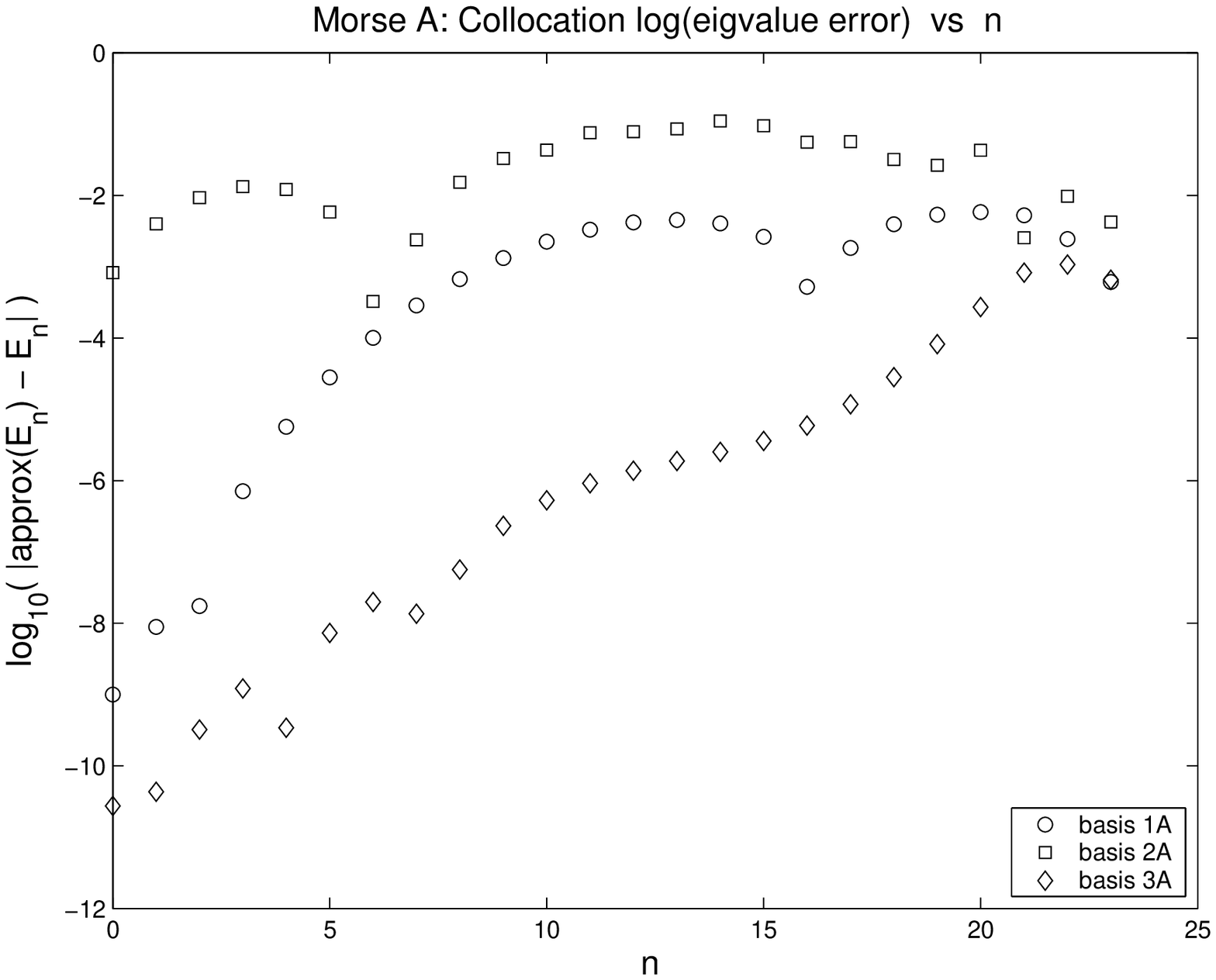}
\includegraphics[width = 2.8in]{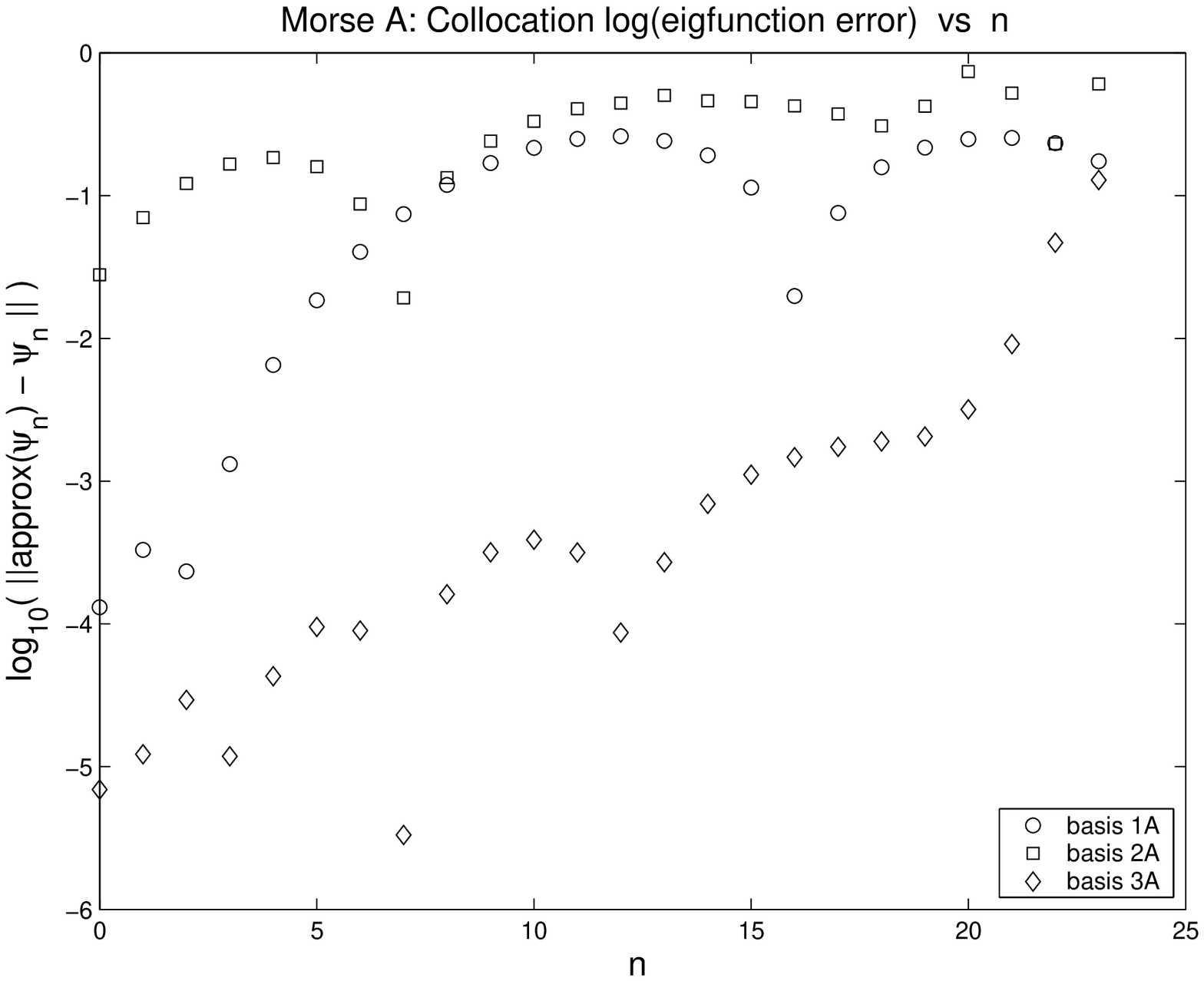}
\caption{Comparing the different basis sets, Morse A: Logarithms of eigenvalue (left column) and eigenfunction (right column) errors vs. index for each discretization method (rows) and each basis
set (circles-set 1A, squares-set 2A, diamonds-set 3A) applied to Morse Hamiltonian A. In each case results were obtained using the global width parameter value which minimizes the mean
eigenfunction error $\epsilon_f$, see table \ref{TabOne} A.
  \label{fig101}}
\end{figure}

\begin{figure}[p]
\centering
\includegraphics[width = 2.8in]{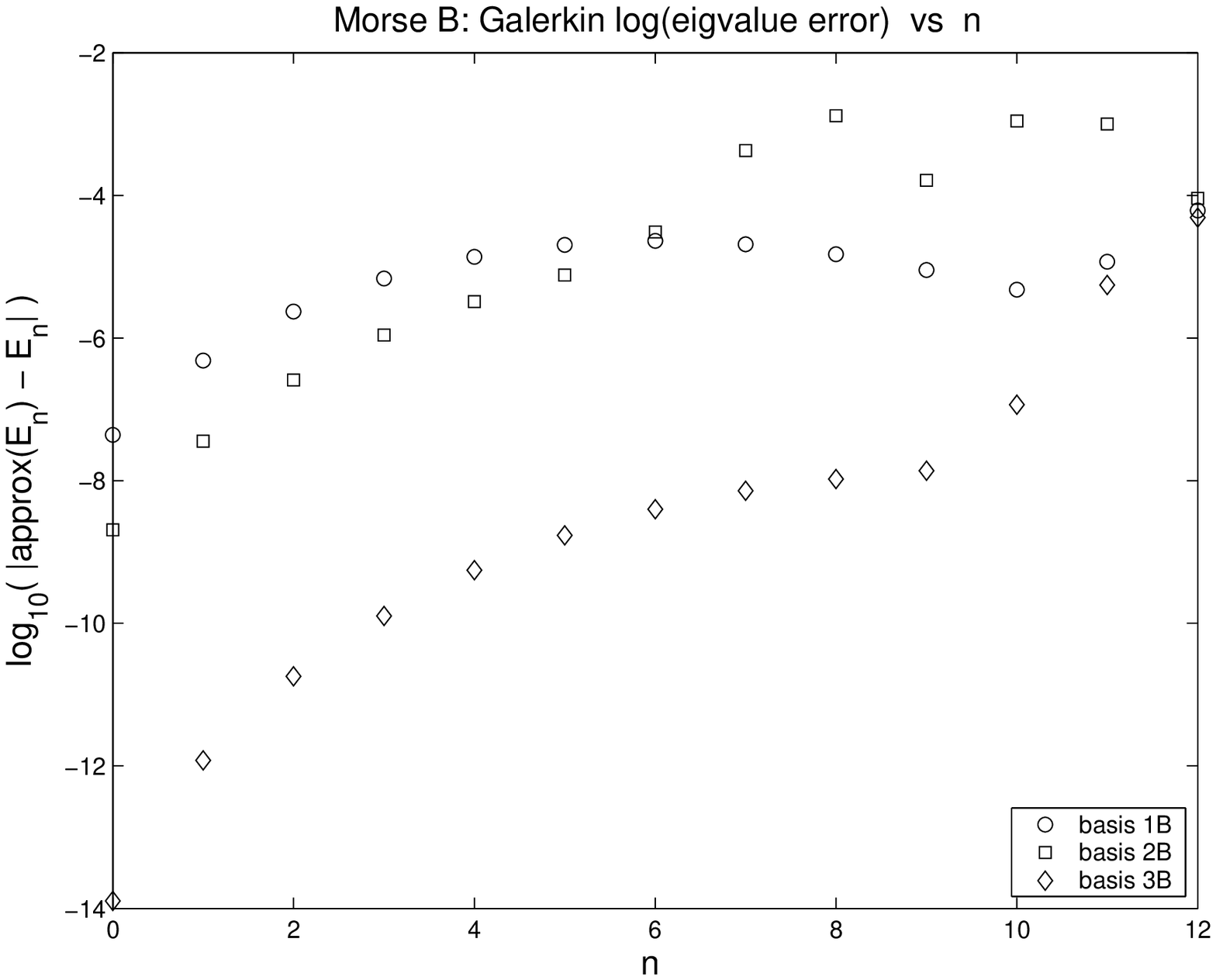}
\includegraphics[width = 2.8in]{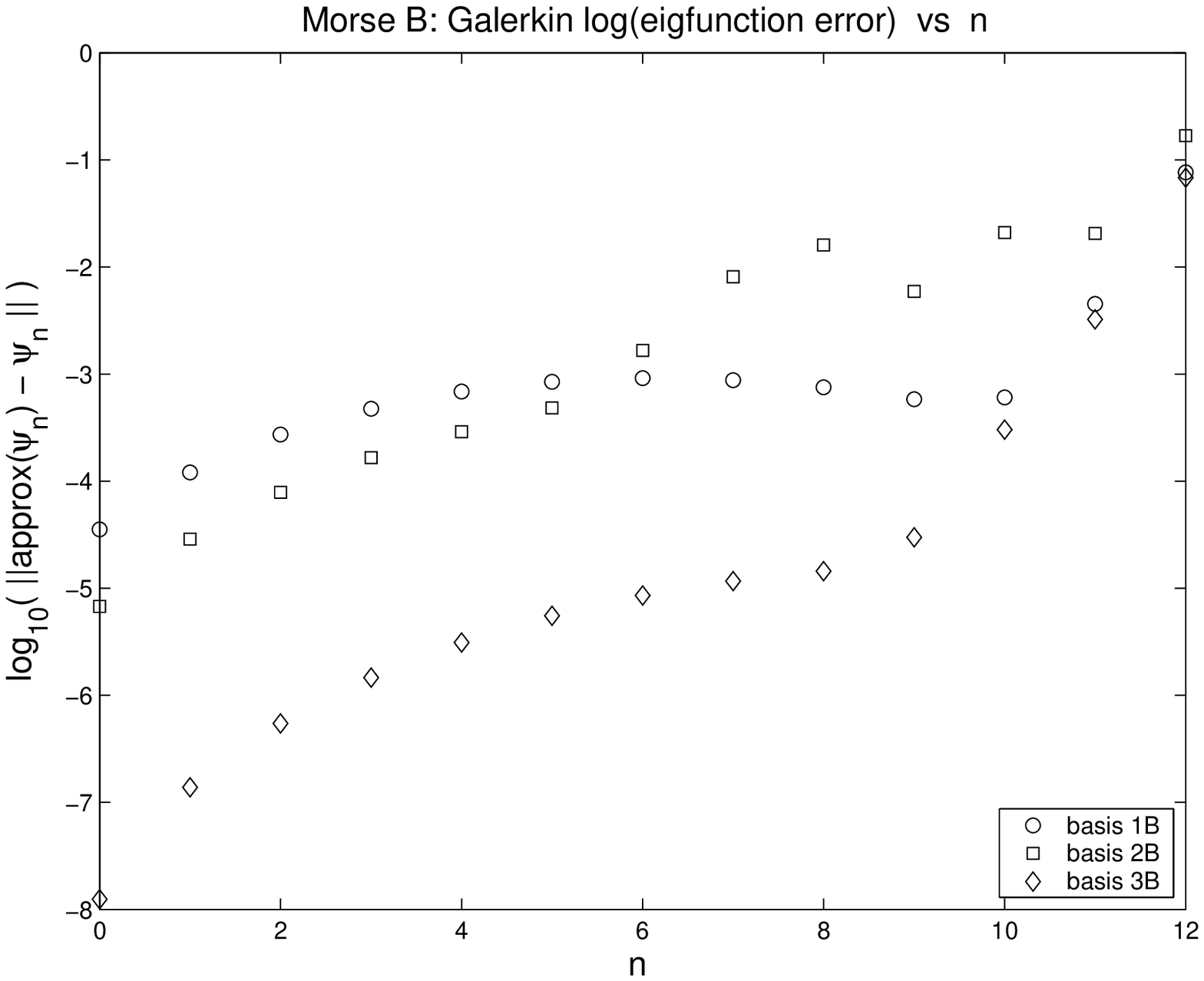}
\includegraphics[width = 2.8in]{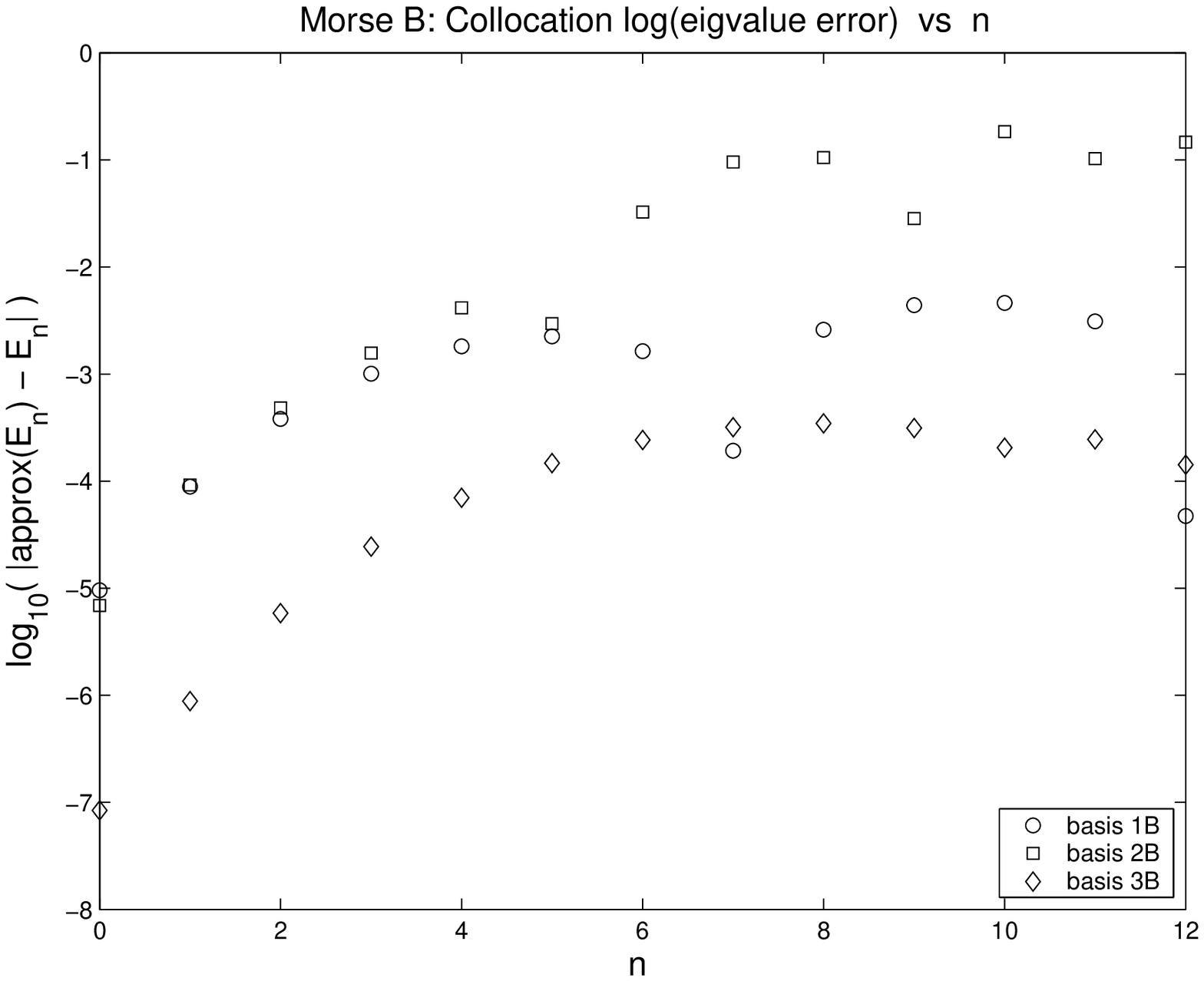}
\includegraphics[width = 2.8in]{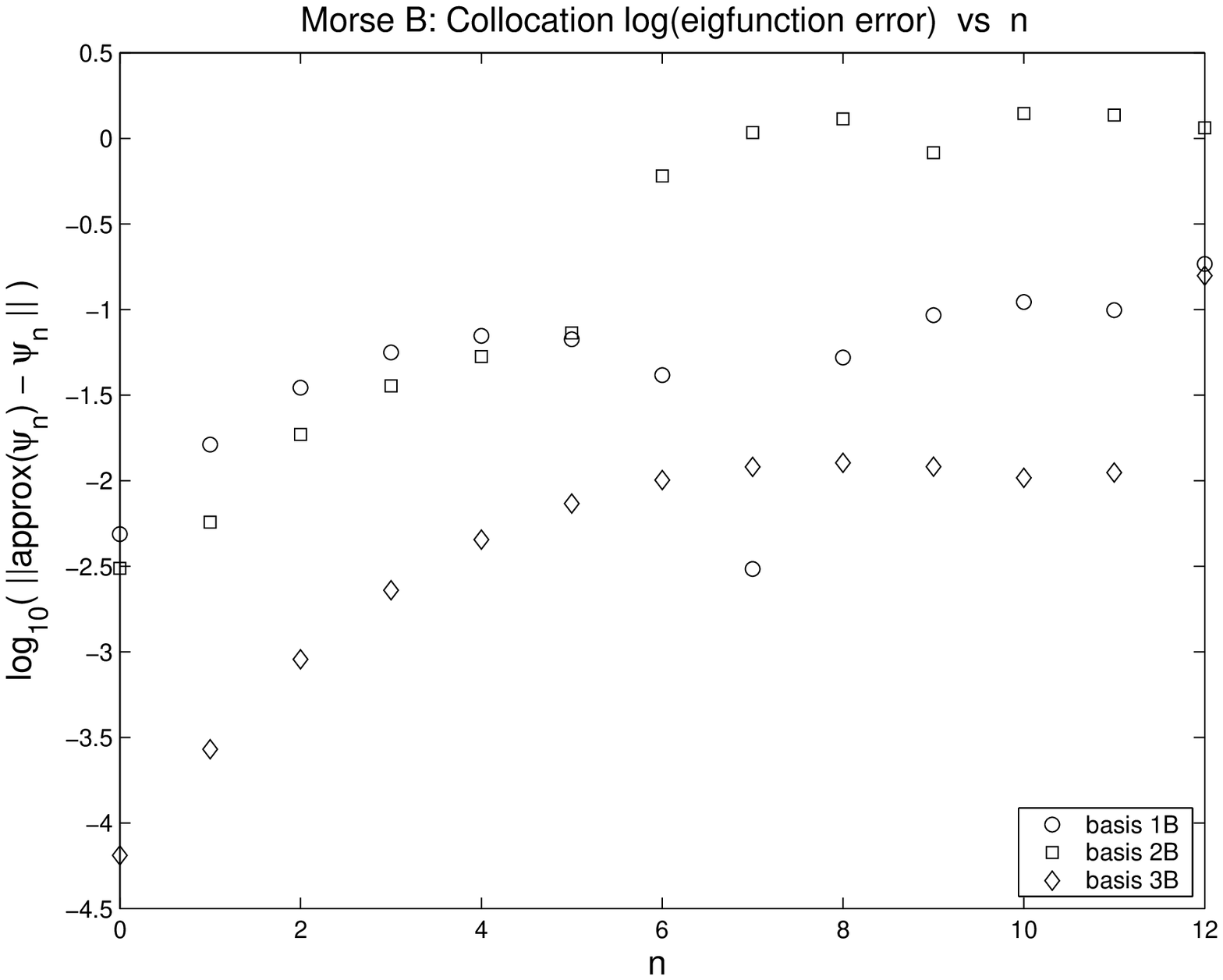}
\caption{ Comparing the different basis sets, Morse B: Logarithms of eigenvalue (left column) and eigenfunction (right column) errors vs. index for each discretization method (rows) and each basis
set (circles-set 1B, squares-set 2B, diamonds-set 3B) applied to Morse Hamiltonian B. In each case (with the exception of basis 2B) results were obtained using the global width parameter value which minimizes the mean
eigenfunction error $\epsilon_f$, see table \ref{TabOne} B.
 \label{fig102}}
\end{figure}

For each basis set the Hamiltonian was discretized using collocation with the $x_i$ serving as collocation nodes, and using Galerkin with analytic calculation of matrix elements.
The average eigenvalue error
\be \epsilon_v = \frac{1}{N+1} \sum_{n=0}^{N} |E_n^a - E_n| \ , \label{EpsV} \ee 
and the average $L_2$ norm of eigenfunction errors
\be \epsilon_f = \frac{1}{N+1} \sum_{n=0}^{N} \| \psi_n^a - \psi_n \|_{L_2} \ , \label{EpsF} \ee 
were calculated for different values of the global width parameter $c$. The superscript $a$ stands for ``approximate'', and $N = 23$ or $12$ for system A or B respectively. Results are shown in figures \ref{fig4}, \ref{fig6}. These figures also illustrate the $c$ dependence of the collocation orthogonalization error
\be \| U^* S U - I \|_{Frobenius} \ ,  \label{CollOrthErr} \ee 
with $U$ being the $48 \times (N+1)$ matrix of eigenvectors calculated by collocation. 
Table \ref{TabOne} 
gives $cond(S)$ and $cond(\Phi)$ at the $c$ values minimizing $\epsilon_f$ for each basis set and discretization method. The only exception is in basis 2B were the minimizing $c$ values gave very ill conditioned bases. Instead larger $c$ was taken so that $cond(S) \approx 10^{12}$ as advocated in \cite{GarLight}.  
For the different discretization methods and bases (with the $c$ values of table \ref{TabOne}) the logarithms of eigenvalue and eigenfunction errors are plotted against their index in figures \ref{fig5}, \ref{fig7}, \ref{fig101}, \ref{fig102}.

Figures \ref{fig101} and \ref{fig102} show that basis sets 3 gave
orders of magnitude better results compared to basis sets 1 and 2, except for a few high
eigenfunctions/eigenvalues where accuracy was similiar. This indicates that 
indeed there is room for improvement in the method of \cite{GarLight} for choosing nodes/widths in hard potential regions. Moreover, the modification suggested here in basis set 3 can produce highly accurate results. 

The results for basis sets 1 and 3 in figures \ref{fig4}, \ref{fig6},
(top and bottom rows) indicate that it may be possible to choose the global width parameter $c$ based on the collocation orthogonalization error. 
The minimum of $\epsilon_f$ for collocation, and for Galerkin, occurs near minima of the collocation ``orth err''. In some cases these minima occur for $c$ values for which
$cond(S)$ is not too large, see table \ref{TabOne}. This supports our notion that it
is desirable to have a criterion for choosing $c$ which is more refined than reaching large $cond(S)$.
 Similiar relations between minima of $\epsilon_f$ and of the collocation orthogonalization error were not observed in basis sets 2 with the quasi randomly
distributed nodes \cite{GarLight}. 


The simplicity of collocation is an important advantage which can save much programing time. Moreover, with a careful choice of basis function parameters collocation can give accurate results. However, figures \ref{fig5}, \ref{fig7} illustrate that in all our calculations the Galerkin results were much
more accurate than those of collocation. 

Examining the bottom row in figures
\ref{fig101}, \ref{fig102}, we see that collocation errors were larger with bases 2A and 2B compared to
bases 1A and 2A. That is, we see that larger errors appeared when using collocation with quasi random DGBs, compared to DGBs whose centers had the same density but were chosen deterministically.



\section{Conclusions and future directions}
\label{Conclude}

In this work several observations about Gaussian radial basis functions
and collocation in Schr\"odinger
eigenproblems were made and tested numerically. It may be hoped that the following conclusions are relevant not only for choosing good Gaussian bases, but also for other types of localized radial basis functions.

In \cite{GarLight} Garashchuk and Light give one of the few examples of efficient basis functions for multi-D Schr\"odinger equations.
An improvement of their approach  may be obtained from our observation that the Gaussians of \cite{GarLight} are too wide in ``hard'' potential regions, $e.g.$ where nuclei are close together in molecular potentials. This led us to suggest choosing nodes/weights using the method of \cite{GarLight}, but with a modified potential function obtained by
replacing steeply rising parts by ``vertical'' walls and flattening nearby
valley regions. We emphasize that our goal is to produce basis
functions which are not overly wide and sparse in hard potential regions.
Using the modified potential above is just one way to achieve this goal,
others are suggested in the future directions below.
Our numerical tests indicate that the changes we propose in the choice
of nodes/widths can give a dramatic improvement in accuracy.

In the literature we often encounter the conception that with appropriately wide basis functions collocation errors are not larger than those of Galerkin \cite{YangPeet1}, \cite{YangPeet2}, \cite{YangPeet3}. However, in our numerical tests the Galerkin accuracy was better than that of collocation, often orders of magnitude better. The difference is probably because we have chosen the optimal global width parameter for each method, while
in \cite{YangPeet1}, \cite{YangPeet2}, \cite{YangPeet3} no special attempt was made to
optimize $c$. Our calculations indicate that caution should be exercised before embracing
collocation as equally accurate to Galerkin.

In \cite{GarLight}, \cite{YangPeet1}, and other places, it is suggested to decrease the global width parameter, making basis functions wider, until the basis is nearly badly conditioned. 
However a more refined method for choosing $c$ is desirable. One supporting fact is that the
optimal $c$ values in table \ref{TabOne} indeed give large $cond(S)$ but not quite as large
as expected in the literature. 
Our numerical tests indicate that the collocation orthogonalization error, defined
in equation (\ref{CollOrthErr}), may guide us in choosing the global width parameter. We found that minima of the collocation orthogonalization
error plotted against the global width parameter occur near minima of eigenfunction/eigenvalue errors calculated by collocation and by the Galerkin approach.

For the future, it may be profitable to attempt a further refinement of the choice of nodes/widths in ``hard'' potential regions. Suppose we have nodes/widths constructed by some distribution method, $e.g.$ that of \cite{GarLight}. Methods
should be explored for identifying nodes lying in hard potential regions, for adding more nodes
to these regions, and for choosing width parameters with respect to potential
values in a neighborhood and not just a point. It seems feasible that these
goals can be achieved for high dimensional systems using only simple local calculations.

Apart from choosing the centers and relative widths of radial basis functions,
a correct scaling of their widths through the global width parameter $c$
is critical for good performance. 
Our suggestion of using the collocation orthogonalization error as a criterion for choosing $c$ may be worth pursuing.
One possibility is to avoid full diagonalization of the collocation Hamiltonian matrix and to measure the orthogonalization error only for the
few lowest, or highest, eigenvectors. An even simpler method may be tried in the
Galerkin approach where the ``variational principle'' holds. In this case the trace of the Hamiltonian matrix 
is minimal for the optimal $c$ (w.r.t. eigenvalue errors). A very simple method for choosing a ``good'' $c$ may possibly be based on this observation.
The work required will not be larger compared
to the method used in \cite{GarLight} based on the condition number of $S^{-1}$.
Some modifications may also be examined. For example, if the matrix elements of $H$ are not calculated exactly we may try to choose $c$ by minimizing the
first few eigenvalues of $S^{-1} H$.

Recall from section \ref{ObsSug} that optimization of functional (\ref{FuncPL}) gives nodes/widths which are consistent with our observations.
So it may be interesting (but probably difficult) to apply the ideas of
Garashchuk and Light from \cite{GarLight} but with the functional (\ref{FuncPL}), rather than
the functional (\ref{FuncGL}). That is,
to attempt identifying a functional form for the densities and width parameters arising from
optimization of functional (\ref{FuncPL}). 

Finally, in addition to further numerical work, a mathematical error analysis for
RBFs applied to Schr\"odingers equation should be attempted.

\vspace{5mm}

\noindent
{\bf Acknowledgements:}  It is a pleasure to thank Yair Goldfarb, David J. Tannor, Antonella Zanna Munthe-Kaas, Hans Z. Munthe-Kaas, Tor S{\o}revik, Jan Petter Hansen,
Lene S{\ae}len, Lars Bojer Madsen, and Jorge Fernandez, for stimulating discussions, pointers to the literature, and encouragement in different stages of this work. This work is part of the computational dynamics of quantum systems (CODY) project, supported by the Norwegian research council.

\begin{table}[h!]
  \noindent
  \begin{tabular}{|c||c|c||c|c||c|c|} \hline
   $ \  $ &
   \multicolumn{2}{c||}{basis set 1A} &
   \multicolumn{2}{c||}{basis set 2A} &
   \multicolumn{2}{c|}{basis set 3A} \\ \hline
   $ \ $ & Galerkin & collocation & Galerkin  & collocation & Galerkin & collocation  \\ \hline 
   $c  $ & $ 0.6   $ & $1.1  $  & 
   $0.8  $ & $1.7  $ &
   $0.6  $ & $0.9  $ \\ \hline
   $cond(\Phi)  $ & $4.07 \times 10^{10}  $ & $ 3.99 \times 10^{3}$& 
   $6.73 \times 10^6  $ & $480  $ &
   $1.2 \times 10^8  $ & $6.53 \times 10^4  $ \\ \hline
   $cond(S)  $ & $4.4 \times 10^9  $ & $8.9 \times 10^4  $ & $6.97 \times 10^8$ & $7.3 \times 10^4$  & $4.76 \times 10^9$ & $9.66 \times 10^6$ \\ \hline \hline 
   $ \  $ &
   \multicolumn{2}{c||}{basis set 1B} &
   \multicolumn{2}{c||}{basis set 2B} &
   \multicolumn{2}{c|}{basis set 3B} \\ \hline
   $ \ $ & Galerkin & collocation & Galerkin  & collocation & Galerkin & collocation \\ \hline 
   $c$ & $1.6$ & $1.7$  & 
   $2$ & $2$  &
   $1.4$ & $1.1$   \\ \hline
   $cond(\Phi)$ & $4.12 \times 10^7$ & $6.71 \times 10^6$ & 
   $1.46 \times 10^{9}  $ & $1.46\times 10^9  $ &
   $5.31\times 10^7$ & $1.02 \times 10^8$ \\ \hline
   $cond(S)  $ & $3.98 \times 10^8$ & $2.21 \times 10^8$  &
   $1.85 \times 10^{12}  $ & $1.85 \times 10^{12}  $ & $4.28 \times 10^9  $ & $1.96 \times 10^{10}  $  \\ \hline
   \end{tabular}
   \caption{The condition numbers of $\Phi$ and $S$ are given at the $c$ (global width parameter) values minimizing the mean eigenfunction error $\epsilon_f$
   (defined in equation (\ref{EpsF})) for each basis set and each discretization method applied to systems A and B. For basis 2B the non optimal value $c=2$ was chosen since the associated $cond(S)$ has the maximal value advocated in \cite{GarLight}. The optimal values were $c=0.9$ for Galerkin, with $cond(S) = 1.68\times 10^{17}$, and $c=1.4$ for collocation, with $cond(S) = 7.79\times 10^{14}$. }
   \label{TabOne}
\end{table}


\pagebreak


\begin{thebibliography}{30}
\bibitem{HighDimCalc} X.G Wang and T. Carrington Jr. {\it A contracted basis-Lanczos  calculation of vibrational levels of methane: Solving the Schr\"{o}dinger equation in nine dimensions,} {\sf J.Chem.Phys. 119 (2003) 101}.
\bibitem{DVRoriginals} D.O. Harris, G.G. Engerholm and W.D. Gwinn, {\it Calculation of matrix elements for one-dimensional quantum mechanical problems and the application to anharmonics oscillators,} {\sf J.Chem.Phys. 43 (1965) 151}; A.S. Dickinson and P.R. Certain,
{\it Calculation of matrix elements for one-dimensional quantum mechanical problems}, {\sf J.Chem.Phys. 49 (1968) 4209;} R.G. Littlejohn, M. Cargo, T.Carrington Jr., K.A. Mitchell, B. Poirier, {\it A general framework for discrete variable representation basis sets,} {\sf J.Chem.Phys. 116 (2002) 8691}.
\bibitem{LightDVR} J.C. Light and T. Carrington Jr. {\it Discrete-Variable Representations and Their Utilization,} {\sf Adv. in Chem.Phys. 114 (2000) 263, ed. I.Prigogine and S.A. Rice.} 
\bibitem{NonProdDVR} I. Degani and D.J. Tannor, {\it Calculating multidimensional discrete variable representations from cubature formulas,} 
{\sf J.Phys.Chem. A 110 (2006) 5395}; M. Cargo. and R.G. Littlejohn. {\it Tetrahedrally invariant discrete variable representation basis on the sphere,} {\sf J.Chem.Phys. 117 (2002) 59}; R. Dawes and T. Carrington Jr. {\it A multidimensional discrete variable representation basis obtained by simultaneous diagonalization,} {\sf J.Chem.Phys. 121 (2004) 726.}
\bibitem{HamLight} I.P. Hamilton and J.C. Light, {\it On distributed Gaussian bases for simple model multidimensional vibrational problems,} {\sf J.Chem.Phys. 84 (1986) 306.}
\bibitem{PoirLight} B. Poirier and J.C. Light, {\it Efficient distributed Gaussian
basis for rovibrational spectroscopy calculations,} {\sf J.Chem.Phys. 113 (2000) 211.}
\bibitem{GarLight} S. Garashchuk and J.C. Light, {\it Quasirandom distributed Gaussian bases for bound problems,} {\sf J.Chem.Phys. 114 (2001) 3929.}
\bibitem{BSplines} B.W. Shore, {\it Comparison of matrix methods applied to the radial Schr\"odinger eigenvalue equation: The Morse potential},
{\sf J.Chem.Phys. 59 (1973) 6450;} B.W. Shore, {\it B-spline expansion bases for excited states and discretized scattering states},
{\sf J.Chem.Phys. 63 (1975) 3385;} H. Bachau, E. Cormier, P. Decleva,
J.E. Hansen, and F. Martin, {\it Applications of B-splines in atomic and
molecular physics,} {\sf Rep.Prog.Phys. 64 (2001) 1815.}
\bibitem{MacDonald} J.K.L. MacDonald, {\it Succesive approximations by the
Rayliegh-Ritz variational method}, {\sf Phs.Rev 43 (1933) 830. }
\bibitem{Arfken} G.B. Arfken and H.J. Weber, {\it Mathematical methods for
physicists,} {\sf Harcourt academic press, fifth ed. (2001) 817.} 
\bibitem{RabitzGIMQ} X.G. Hu, T.S. Ho, and H. Rabitz, {\it The collocation method based on a generalized inverse multiquadric basis for bound state problems,}
{\sf Comp.Phys.Comm. 113 (1998) 168.}
\bibitem{SobolSoft} W. Putsch\"ogl, {\it Interface to Sobol GSL implementation,} {\sf the matlab central file exchange (2006)  http://www.mathworks.com/matlabcentral/}
\bibitem{YangPeet1} W. Yang and A.C. Peet, {\it The collocation method for bound solutions of the Schr\"odinger equation,} {\sf Chem.Phys.Lett. 153 (1988) 98}. 
\bibitem{YangPeet2} W. Yang and A.C. Peet, {\it The collocation method for 
calculating vibrational bound states of molecular systems with application to 
Ar-HCl,} {J.Chem.Phys. 90 (1989) 1746.}
\bibitem{YangPeet3} W. Yang and A.C. Peet, {\it A method for calculating vibrational bound states: iterative solution of the collocation equations constructed from localized basis sets,} {J.Chem.Phys. 92 (1990) 522.}
\bibitem{InitialCollRefs} L. Collatz, {\it The numerical treatment of
differential equations,} {\sf Springer Verlag, third ed. (1960).}
\bibitem{KansaEtc} E.J. Kansa, {\it Multiquadrics - a scattered data approximation scheme with applications to computational fluid dynamics - I,} {\sf Computers Math.Applic. 19 (1990) 127;} E.J. Kansa, {\it Multiquadrics - a scattered data approximation scheme with applications to computational fluid dynamics - II,} {\sf Computers Math.Applic. 19 (1990) 147.}
\bibitem{MorseHamAnalytic} M.M. Nieto and L.M. Simmons Jr. {\it Eigenstates, coherent states, and uncertainty
products of the Morse oscillator,} {\sf Phys.Rev.A 19 (1979) 438.}
\bibitem{MathRadBFuncs} M.D. Buhmann, {\it Radial basis functions,} {\sf Acta
Numerica (2000) 1}; R. Schaback and H. Wendland, {\it Kernel techniques: From
machine learning to meshless methods,} {\sf Acta Numerica (2006) 543.}
\end{thebibliography}
\end{document}